\documentclass[twocolumn,prc,preprintnumbers,superscriptaddress,showpacs,amsmath,amssymb,floatfix]{revtex4}
\usepackage{graphicx}
\usepackage{dcolumn}
\usepackage{amsmath}
\usepackage{colordvi}
\usepackage{url}
\usepackage{longtable}
\usepackage{psfig}
\begin{document}
\title{Measurement of the generalized form factors \\
near threshold via $\gamma^* p \to n\pi^+$ at high $Q^2$}

\date{\today}

\newcommand*{\ANL}{Argonne National Laboratory, Argonne, Illinois 60439}
\newcommand*{\ANLindex}{1}
\affiliation{\ANL}
\newcommand*{\ASU}{Arizona State University, Tempe, Arizona 85287-1504}
\newcommand*{\ASUindex}{2}
\affiliation{\ASU}
\newcommand*{\UCLA}{University of California Los Angeles, Los Angeles, California  90095-1547}
\newcommand*{\UCLAindex}{3}
\affiliation{\UCLA}
\newcommand*{\CSUDH}{California State University, Dominguez Hills, Carson, CA 90747}
\newcommand*{\CSUDHindex}{4}
\affiliation{\CSUDH}
\newcommand*{\CANISIUS}{Canisius College, Buffalo, NY}
\newcommand*{\CANISIUSindex}{5}
\affiliation{\CANISIUS}
\newcommand*{\CMU}{Carnegie Mellon University, Pittsburgh, Pennsylvania 15213}
\newcommand*{\CMUindex}{6}
\affiliation{\CMU}
\newcommand*{\CUA}{Catholic University of America, Washington, D.C. 20064}
\newcommand*{\CUAindex}{7}
\affiliation{\CUA}
\newcommand*{\SACLAY}{CEA, Centre de Saclay, Irfu/Service de Physique Nucl\'eaire, 91191 Gif-sur-Yvette, France}
\newcommand*{\SACLAYindex}{8}
\affiliation{\SACLAY}
\newcommand*{\CNU}{Christopher Newport University, Newport News, Virginia 23606}
\newcommand*{\CNUindex}{9}
\affiliation{\CNU}
\newcommand*{\UCONN}{University of Connecticut, Storrs, Connecticut 06269}
\newcommand*{\UCONNindex}{10}
\affiliation{\UCONN}
\newcommand*{\EDINBURGH}{Edinburgh University, Edinburgh EH9 3JZ, United Kingdom}
\newcommand*{\EDINBURGHindex}{11}
\affiliation{\EDINBURGH}
\newcommand*{\FU}{Fairfield University, Fairfield CT 06824}
\newcommand*{\FUindex}{12}
\affiliation{\FU}
\newcommand*{\FIU}{Florida International University, Miami, Florida 33199}
\newcommand*{\FIUindex}{13}
\affiliation{\FIU}
\newcommand*{\FSU}{Florida State University, Tallahassee, Florida 32306}
\newcommand*{\FSUindex}{14}
\affiliation{\FSU}
\newcommand*{\Genova}{Universit$\grave{a}$ di Genova, 16146 Genova, Italy}
\newcommand*{\Genovaindex}{15}
\affiliation{\Genova}
\newcommand*{\GWUI}{The George Washington University, Washington, DC 20052}
\newcommand*{\GWUIindex}{16}
\affiliation{\GWUI}
\newcommand*{\ISU}{Idaho State University, Pocatello, Idaho 83209}
\newcommand*{\ISUindex}{17}
\affiliation{\ISU}
\newcommand*{\INFNFE}{INFN, Sezione di Ferrara, 44100 Ferrara, Italy}
\newcommand*{\INFNFEindex}{18}
\affiliation{\INFNFE}
\newcommand*{\INFNFR}{INFN, Laboratori Nazionali di Frascati, 00044 Frascati, Italy}
\newcommand*{\INFNFRindex}{19}
\affiliation{\INFNFR}
\newcommand*{\INFNGE}{INFN, Sezione di Genova, 16146 Genova, Italy}
\newcommand*{\INFNGEindex}{20}
\affiliation{\INFNGE}
\newcommand*{\INFNRO}{INFN, Sezione di Roma Tor Vergata, 00133 Rome, Italy}
\newcommand*{\INFNROindex}{21}
\affiliation{\INFNRO}
\newcommand*{\ORSAY}{Institut de Physique Nucl\'eaire ORSAY, Orsay, France}
\newcommand*{\ORSAYindex}{22}
\affiliation{\ORSAY}
\newcommand*{\ITEP}{Institute of Theoretical and Experimental Physics, Moscow, 117259, Russia}
\newcommand*{\ITEPindex}{23}
\affiliation{\ITEP}
\newcommand*{\JMU}{James Madison University, Harrisonburg, Virginia 22807}
\newcommand*{\JMUindex}{24}
\affiliation{\JMU}
\newcommand*{\KNU}{Kyungpook National University, Daegu 702-701, Republic of Korea}
\newcommand*{\KNUindex}{25}
\affiliation{\KNU}
\newcommand*{\LPSC}{LPSC, Universite Joseph Fourier, CNRS/IN2P3, INPG, Grenoble, France
}
\newcommand*{\LPSCindex}{26}
\affiliation{\LPSC}
\newcommand*{\UNH}{University of New Hampshire, Durham, New Hampshire 03824-3568}
\newcommand*{\UNHindex}{27}
\affiliation{\UNH}
\newcommand*{\NSU}{Norfolk State University, Norfolk, Virginia 23504}
\newcommand*{\NSUindex}{28}
\affiliation{\NSU}
\newcommand*{\OHIOU}{Ohio University, Athens, Ohio  45701}
\newcommand*{\OHIOUindex}{29}
\affiliation{\OHIOU}
\newcommand*{\ODU}{Old Dominion University, Norfolk, Virginia 23529}
\newcommand*{\ODUindex}{30}
\affiliation{\ODU}
\newcommand*{\RPI}{Rensselaer Polytechnic Institute, Troy, New York 12180-3590}
\newcommand*{\RPIindex}{31}
\affiliation{\RPI}
\newcommand*{\URICH}{University of Richmond, Richmond, Virginia 23173}
\newcommand*{\URICHindex}{32}
\affiliation{\URICH}
\newcommand*{\ROMAII}{Universita' di Roma Tor Vergata, 00133 Rome Italy}
\newcommand*{\ROMAIIindex}{33}
\affiliation{\ROMAII}
\newcommand*{\MSU}{Skobeltsyn Nuclear Physics Institute, Skobeltsyn Nuclear Physics Institute, 119899 Moscow, Russia}
\newcommand*{\MSUindex}{34}
\affiliation{\MSU}
\newcommand*{\SCAROLINA}{University of South Carolina, Columbia, South Carolina 29208}
\newcommand*{\SCAROLINAindex}{35}
\affiliation{\SCAROLINA}
\newcommand*{\JLAB}{Thomas Jefferson National Accelerator Facility, Newport News, Virginia 23606}
\newcommand*{\JLABindex}{36}
\affiliation{\JLAB}
\newcommand*{\UTFSM}{Universidad T\'{e}cnica Federico Santa Mar\'{i}a, Casilla 110-V Valpara\'{i}so, Chile}
\newcommand*{\UTFSMindex}{37}
\affiliation{\UTFSM}
\newcommand*{\GLASGOW}{University of Glasgow, Glasgow G12 8QQ, United Kingdom}
\newcommand*{\GLASGOWindex}{38}
\affiliation{\GLASGOW}
\newcommand*{\VT}{Virginia Polytechnic Institute and State University, Blacksburg, Virginia   24061-0435}
\newcommand*{\VTindex}{39}
\affiliation{\VT}
\newcommand*{\VIRGINIA}{University of Virginia, Charlottesville, Virginia 22901}
\newcommand*{\VIRGINIAindex}{40}
\affiliation{\VIRGINIA}
\newcommand*{\WM}{College of William and Mary, Williamsburg, Virginia 23187-8795}
\newcommand*{\WMindex}{41}
\affiliation{\WM}
\newcommand*{\YEREVAN}{Yerevan Physics Institute, 375036 Yerevan, Armenia}
\newcommand*{\YEREVANindex}{42}
\affiliation{\YEREVAN}

\newcommand*{\NOWMSU}{Skobeltsyn Nuclear Physics Institute, Skobeltsyn Nuclear Physics Institute, 119899 Moscow, Russia}
\newcommand*{\NOWINFNFR}{INFN, Laboratori Nazionali di Frascati, 00044 Frascati, Italy}
\newcommand*{\NOWINFNGE}{INFN, Sezione di Genova, 16146 Genova, Italy}
\newcommand*{\NOWUCONN}{University of Connecticut, Storrs, Connecticut 06269}

\author {K.~Park} 
\affiliation{\JLAB}
\affiliation{\SCAROLINA}
\author {R.W.~Gothe} 
\affiliation{\SCAROLINA}

\author {K.P. ~Adhikari} 
\affiliation{\ODU}
\author {D.~Adikaram} 
\affiliation{\ODU}
\author {M.~Anghinolfi} 
\affiliation{\INFNGE}
\author {H.~Baghdasaryan} 
\affiliation{\VIRGINIA}
\affiliation{\YEREVAN}
\author {J.~Ball} 
\affiliation{\SACLAY}
\author {M.~Battaglieri} 
\affiliation{\INFNGE}
\author {V.~Batourine} 
\affiliation{\JLAB}
\author {I.~Bedlinskiy} 
\affiliation{\ITEP}
% ...

\collaboration{The CLAS Collaboration}
\noaffiliation


\begin{abstract}
We report the first extraction of the pion-nucleon multipoles near the 
production threshold for the $n\pi^+$ channel at relatively high momentum transfer 
($Q^2$ up to 4.2 $\rm{GeV^2}$). The dominance of the $s$-wave transverse multipole ($E_{0+}$), 
expected in this region, allowed us to access the generalized form factor $G_1$
within the light-cone sum rule (LCSR) framework as well as the axial
form factor $G_A$. The data analyzed in this work were collected by the nearly 
$4\pi$ CEBAF Large Acceptance Spectrometer (CLAS) using a 5.754 $\rm{GeV}$ electron beam 
on a proton target. The differential cross section and the $\pi-N$-multipole 
$E_{0+}/G_D$ were measured using two different methods, the LCSR and a direct 
multipole fit. The results from the two methods are found to be consistent and 
almost $Q^2$ independent.
\end{abstract}

\maketitle


\section{Introduction}
 Pion threshold photo- and electroproduction has a long history with continuous interest from both experimental and theoretical sides. These studies are of interest  because the vanishing pion mass approximation in chiral symmetry, supplemented by current algebra, allows exact predictions to be made for the threshold cross sections, so-called Low-Energy Theorems (LET)~\cite{NMKnoll}\cite{YNambu1}\cite{YNambu2}. As a prominent example, the LET establish a connection between charged pion electroproduction and the axial form factor of the nucleon. In the real world, the finite pion mass cannot be ignored ($m_{\pi}/m_N \sim 1/7$). The study of finite pion mass corrections to LET was a topical field in high energy physics in the late sixties and early seventies before the discovery of Bjorken scaling in Deep Inelastic Scattering (DIS) and the advent of Quantum Chromodynamics (QCD)~\cite{AIVainshtein}. (A monograph on pion-electroproduction~\cite{EAmaldi} addresses many of these developments.) 

  In the eighties and nineties, a renewed interest in threshold pion production was triggered by the extensive data that became available on $\gamma p \to \pi^0 p$~\cite{EMazzucato}\cite{RBeck} and $\gamma^* p \to \pi^0 p$ at $Q^2 = 0.04-0.1\;\rm{GeV^2}$~\cite{TPWelch} ($q = p_e - p_{e'}, Q^2 = -q^2$). At the same time, the advent of CHiral Perturbation Theory (CHPT) has allowed the systematic expansion of physical low-energy observables in powers of the pion mass and momentum. 
 The new insight brought by CHPT calculations is that certain loop diagrams produce non-analytic contributions to scattering amplitudes that are lost in the naive expansion in the pion mass~\cite{AIVainshtein}\cite{SScherer}. The expansion at small photon virtualities $Q^2$ has to be done with care as the limits $m_{\pi} \to 0$ and $Q^2 \to 0$ do not commute in general~\cite{VBernard}. The LET predictions that include CHPT corrections seem to be in good agreement with experimental data on pion photoproduction~\cite{DDrechsel}. Experimental results on the $s$-wave electroproduction cross section for $Q^2 \sim 0.1\;\rm{GeV^2}$ are also consistent with CHPT calculations when chiral loops are taken into account~\cite{VBernard1}\cite{VBernard2}.

 We report the extraction of the multipole $E_{0+}$ near pion threshold in the charged single pion electroproduction channel (${e}p \to e'n\pi^+$) with a nearly 6 GeV electron beam incident on a proton target. This experimental data set allowed us to study near-threshold pion production at photon virtualities $Q^2$ up to $\sim4.2\;\rm{GeV^2}$. This experiment is a major step forward and requires very good energy resolution in order to approach the pion production threshold, where the $p$-wave contribution of the $M_{1+}$ multipole is suppressed.\\

\section{Light Cone Sum Rule model}
 In the traditional derivation of LET using the Partially Conserved Axial Current (PCAC) approximation and current algebra, $Q^2$ is not assumed to be small and the expansion in powers of the pion mass involves two parameters: $m_{\pi}/m_N$ and $m_{\pi}Q^2/m^3_N$ ~\cite{AIVainshtein}\cite{SScherer}. At high $Q^2$, the second parameter can still be kept small but, in this case, the pion is not soft in the target rest frame, even though at threshold it is soft in the $\pi-\rm{N}$ final state center-of-mass frame. For the threshold kinematics, this affects in particular the contribution of pion emission from the initial state \cite{PVPobylitsa}. The LET is formally valid (modulo CHPT loop corrections~\cite{VBernard}) for momentum transfers as large as $Q^2 \sim m^2_N$. However, no dedicated experimental study of threshold pion production in the $Q^2 \sim 1\;\rm{GeV^2}$ region has been carried out so far. For $m_{\pi}Q^2/m^3_N = O(1)$, the LET breaks down: the initial state pion radiation occurs at time scales of the order $1/m_N$ rather than $1/m_{\pi}$, requiring additional contributions of hadronic intermediate states other than the nucleon. Finally, at very large momentum transfers ($Q^2 \gg 1\;\rm{GeV^2}$), one can factorize hard-scale contributions as coefficient functions in front of soft contributions involving small momenta, allowing the use of current algebra (or CHPT) for the latter, but not for the amplitude as a whole~\cite{PVPobylitsa}.

 For asymptotically large $Q^2$, the standard perturbative QCD (pQCD) collinear factorization technique~\cite{AVEfremov}\cite{GPLepage} becomes applicable, and the helicity-conserving $E_{0+}$ multipoles can be calculated (at least for $m_{\pi} = 0$) in terms of chirally rotated nucleon distribution amplitudes. Resolving the onset of the pQCD regime at very large momenta is difficult because of the competition between the factorized contribution, proportional to $(\alpha_s(Q)/2\pi)^2$ (which runs with $Q^2$), and the non-perturbative soft contributions. The latter are nominally suppressed by extra powers of $Q^2$ but are susceptible to end-point contributions that cause them to contribute even at very high $Q^2$.  

 The light-cone sum rule (LCSR) approach has addressed this problem, and a realistic QCD-motivated model for the $Q^2$ dependence of both the transverse $E_{0+}$ and the longitudinal $L_{0+}$ $s$-wave multipoles has been developed~\cite{VBraun} for the $Q^2 \sim 1 - 10 \;\rm{GeV^2}$ near-threshold region that can now be accessed by the presented experimental data. A technique was developed~\cite{VMBraun00} to calculate baryon form factors for moderately large $Q^2$ using LCSR~\cite{VLChernyak}\cite{IIBalitsky}. The same technique~\cite{VMBraun00} is applied to pion electroproduction.  This approach is attractive because in LCSR, soft contributions to the form factors are calculated in terms of the same nucleon distribution amplitudes that enter the pQCD calculation without double counting. Thus, the LCSR provide the most direct relation of the hadron form factors and nucleon distribution amplitudes that is currently available, without using other nonperturbative parameters. 

 The relevant generalized form factors were estimated in the LCSR approach~\cite{VMBraun01} for the range of momentum transfers $Q^2 \sim 5 - 10\;\rm{GeV^2}$. For this work, the sum rules have been re-derived in~\cite{VMBraun01}, taking into account the semi-disconnected pion-nucleon contributions in the intermediate state. The applicability of the sum rules can be extended to the lower $Q^2$ region, and the LET are indeed reproduced at $Q^2 \sim 1\;\rm{GeV^2}$ to the required accuracy $O(m_{\pi})$.  The results presented here essentially interpolate between the large $Q^2$ limit considered in~\cite{VMBraun01} and the standard LET predictions at low momentum transfers. Two generalized form factors are introduced that describe the $Q^2$-dependence of the $s$-wave multipoles of pion electroproduction at the threshold kinematics. In a simplified approach, the LCSR correlation function for the electroproduction close to threshold shows dominance of $s$-wave form-factor-like contributions.\\

\section{The generalized form factors $G_1$ and $G_2$ from LCSR}\label{gff_lcsr}
\subsection{Differential cross sections and form factors}
 In the one-photon-exchange approximation, the single pion electroproduction cross section factorizes as
\begin{eqnarray} 
\frac{d^4\sigma}{dQ^2dWd\Omega_\pi^*} = |J| \Gamma_v \frac{d^2\sigma_u}{d\Omega_\pi^*} ~,
\end{eqnarray}
where
\begin{eqnarray} 
|J|\Gamma_v&=&\frac{\alpha}{2\pi^2Q^2}\frac{(W^2-M_p^2)E_f}{2M_pE_i(1-\epsilon)} ~,\ \nonumber\\
\epsilon&=&\big[1+2\big(1+\frac{\nu^2}{Q^2}\big)\tan^2\frac{\theta_e}{2}\big]^{-1}, \nonumber
\end{eqnarray}
and
\begin{eqnarray} 
\frac{d^2\sigma_u}{d\Omega_\pi^*} &=&\sigma_T+\epsilon\sigma_L + \epsilon\sigma_{TT} \cos2\phi_{\pi}^* \nonumber\\
 & &+ {\sqrt{2\epsilon(1+\epsilon)}}\sigma_{LT}\cos\phi_{\pi}^*  ~.  \nonumber
\end{eqnarray}
 The parameter $\epsilon$ represents the virtual photon polarization and $\Gamma_v$ is the flux of virtual photons. $E_i$ and $E_f$ are energies of the initial and scattered electrons respectively. The angle $\phi^*_{\pi}$ is the azimuthal rotation of the $n\pi^+$ plane with respect to the electron scattering plane $(e,e')$, $\nu$ (=$E_i -E_f$) is the energy transfer of electron, $\theta_e$ is the polar angle of the scattered electron in the Lab system, $\Omega_\pi^*$ is the solid angle of pion in the center-of-mass frame, and $W$ is the invariant mass. In the absence of a transverse polarization of the target nucleon, the cross section does not depend on $\phi_e$. For an electron beam and a proton target, the center-of-mass differential cross section $d^2\sigma_u$ depends on the virtual photon polarization ($\epsilon$) through four structure functions : $\sigma_T+\epsilon\sigma_L$ and the interference terms $\sigma_{TT}$ and $\sigma_{LT}$.  Four structure functions are determined by a fit to the $\phi^*_{\pi}$-dependent differential cross section. The partial wave decomposition by Legendre polynomials of the structure functions in the limit of angular momenta $l \le 2$ is given by~\cite{VBraun}
\begin{eqnarray}\label{eq:stf_legen}
&&\sigma_{T} + \epsilon \sigma_L = \sum_{l=0}^{n} D_l^{T+L} P_l (\cos\theta_\pi^*)~,\nonumber \\
&&\sigma_{TT} =  \sin^2\theta_\pi^* \sum_{l=0}^{n-2}D^{TT}_l P_l (\cos\theta_\pi^*)\;,~\rm{and}\nonumber\\
&&\sigma_{LT} = \sin\theta_\pi^* \sum_{l=0}^{n-1}D_l^{LT}P_l (\cos\theta_\pi^*)~, 
\end{eqnarray}
 where the coefficients ($D^{T+L}_0$, $D^{T+L}_1$, $D^{T+L}_2$, $D^{TT}_0$, $D^{LT}_0$ and $D^{LT}_1$) depend on seven complex multipoles since the Legendre coefficients are directly related to multipole decomposition. The quantity $\theta_\pi^*$ is  the $\pi^+$ polar angle in the center-of-mass frame. In the LCSR approach in the pion threshold region with vanishing pion mass, the Legendre coefficients can be described in terms of the generalized form factors by
\begin{widetext}
\begin{eqnarray}\label{eq:coeff1}
D^{T+L}_0 &=& \frac{1}{f^2_\pi} \biggr[ \frac{4 \vec{k_i}^2 Q^2}{m_N^2} |G_1^{n\pi^+}|^2 + \frac{{c_\pi}^2 {g_A}^2 \vec{k_f}^2}{W^2 - m_N^2} Q^2 m_N^2 {G^n_M}^2 + \epsilon \left( \vec{k_i}^2 |G_2^{n\pi^+}|^2 +   \frac{4{c_\pi}^2 {g_A}^2 \vec{k_f}^2}{W^2 - m_N^2}m_N^4 {G^n_E}^2 \right) \biggr]~,  \nonumber\\ 
\nonumber\\
D^{T+L}_1 &=& \frac{1}{f^2_\pi} \frac{4{c}_\pi {g_A} |k_i||k_f|}{W^2 - m_N^2} \biggr( Q^2 {G^n_M} Re(G_1^{n\pi^+}) -\epsilon m_N^2 {G^n_E} Re(G_2^{n\pi^+}) \biggr)~, \;\rm{and\nonumber}\\ 
D^{LT}_0 &=& -\frac{1}{f^2_\pi} \frac{{c}_\pi {g_A} |k_i||k_f|}{W^2 - m_N^2} Q m_N \biggr( {G^n_M} Re(G_2^{n\pi^+}) +4 {G^n_E} Re(G_1^{n\pi^+}) \biggr)~, 
\end{eqnarray}
\end{widetext}
 where $G^n_M$ and $G^n_E$ are the magnetic and electric Sachs form factors of the neutron (due to pion emission off the initial proton), $c_\pi = \sqrt{2}$ is the isospin factor, $f_\pi = 93$ MeV is the pion decay constant, and $g_A = 1.267$ is the axial coupling. For charged pion production additional contributions, $G_{1,2}^{n,\pi^+}$ arise from the chiral rotation of the electromagnetic current. These are not present for neutral pion production. The Legendre moment $D_{TT}$ is zero since $d$-waves are absent. Parametrizations developed previously~\cite{PEBosted} for the electric and magnetic neutron form factors ($G_M^n$ and $G_E^n$, respectively) are used in this analysis.
\subsection{$G_E^n$ dependence on $G_1$ and $G_2$}
As described above, experimental data for $G_E^n$ at high momentum transfers are lacking. The quality of predictions for $G_E^n$ at high momentum transfers also remains poor. Thus, the generalized form factor will be extracted here under the assumption of $m_{\pi}\sim0$ with and without taking $G_E^n$ into account. At first, if we take $G_E^n \sim 0$, Eq.~(\ref{eq:coeff1}) can be re-written as
\begin{widetext}
\begin{eqnarray}\label{eq:coeff2}
D^{T+L}_0 &=& \frac{1}{f^2_\pi} \biggr[ \frac{4 \vec{k_i}^2 Q^2}{m_N^2} |G_1^{n\pi^+}|^2 + \frac{{{c_\pi}^2} {g_A}^2 \vec{k_f}^2}{W^2 - m_N^2} Q^2 m_N^2 {G^n_M}^2 + \epsilon \left( \vec{k_i}^2 |G_2^{n\pi^+}|^2  \right) \biggr]~, \nonumber\\ 
\nonumber\\
D^{T+L}_1 &=& \frac{1}{f^2_\pi} \frac{4{c}_\pi {g_A} |k_i||k_f|}{W^2 - m_N^2} \biggr( Q^2 G^n_M Re(G_1^{n\pi^+}) \biggr) ~,~\rm{and}\nonumber\\
D^{LT}_0 &=& -\frac{1}{f^2_\pi} \frac{{c}_\pi {g_A} |k_i||k_f|}{W^2 - m_N^2} Q m_N \biggr( G^n_M Re(G_2^{n\pi^+})  \biggr) ~.
\end{eqnarray}
\end{widetext}
 Here $k_i$, $k_f$ are the center-of-mass momenta in the initial and final states, respectively~\cite{VBraun}. The low-energy theorems (LET) relate the $s$-wave multipoles, or equivalently the form factors $G_1$ and $G_2$ at the pion threshold, to the nucleon electromagnetic and axial form factors for the $n\pi^+$ channel: 
\begin{eqnarray}\label{eq:ff0}
\frac{Q^2}{m_N^2} G_1^{n\pi^+} &=& \frac{g_A}{\sqrt{2}}\frac{Q^2}{Q^2+2m_N^2}G_M^n + \frac{1}{\sqrt{2}}G_A ~\nonumber\\
G_2^{n\pi^+} &=& \frac{2\sqrt{2}g_Am_N^2}{Q^2+2m_N^2}G_E^n~
\end{eqnarray}
 since $G_E^n \sim 0$ in Eq.~(\ref{eq:ff0}), $G_2^{n\pi^+}$ is negligible. Therefore, only two terms survive from Eq.~(\ref{eq:coeff2}), since in the $n\pi^+$ channel $D^{LT}_0=0$ (being $G_E^n = 0$) and $D^{TT}_0=0$ (due to absence of $d$-waves).
\begin{eqnarray}\label{eq:coeff3}
D^{T+L}_0 &=& \frac{1}{f^2_\pi} \biggr[ \frac{4 \vec{k_i}^2 Q^2}{m_N^2} |G_1^{n\pi^+}|^2 + \frac{{c_\pi}^2 {g_A}^2 \vec{k_f}^2}{W^2 - m_N^2} Q^2 m_N^2 {G^n_M}^2 \biggr] \nonumber\\ 
\nonumber\\
D^{T+L}_1 &=& \frac{1}{f^2_\pi} \frac{4{c}_\pi {g_A} |k_i||k_f|}{W^2 - m_N^2} \biggr( Q^2 {G^n_M} Re(G_1^{n\pi^+}) \biggr) .\nonumber
\end{eqnarray}
In this case, we can extract the $E^{n\pi^+}_{0+}$ amplitude by using its relation to the form factor ($G_1^{n\pi^+}$)~\cite{VBraun} both normalized by the dipole form factor ($G_D = 1/(1+Q^2/\mu_0)^2$, $\mu_0=0.71$):
\begin{eqnarray}\label{eq:ff1}
\frac{E_{0+}^{n\pi^+}}{G_D} = \frac{\sqrt{4\pi\alpha_{em}}}{8\pi}\frac{Q^2\sqrt{Q^2+4m_p^2}}{m_p^3 f_\pi}\frac{G_1^{n\pi^+}}{G_D}~.
\end{eqnarray}
%
%
%
%
%
Alternatively, we could take a non-zero value of $G_E^n$ into account, but there are no constraints on the imaginary parts of $G_1$ and $G_2$, since only electron-helicity independent data are available. The real parts of $G_1$ and $G_2$ can still be determined from $D^{T+L}_1$ and $D^{LT}_0$, but to solve for the imaginary parts of $G_1$ and $G_2$, further assumptions are required.

\section{Experiment} 
\begin{figure}[!htb]
\begin{center}
	\includegraphics[angle=0,width=0.49\textwidth]{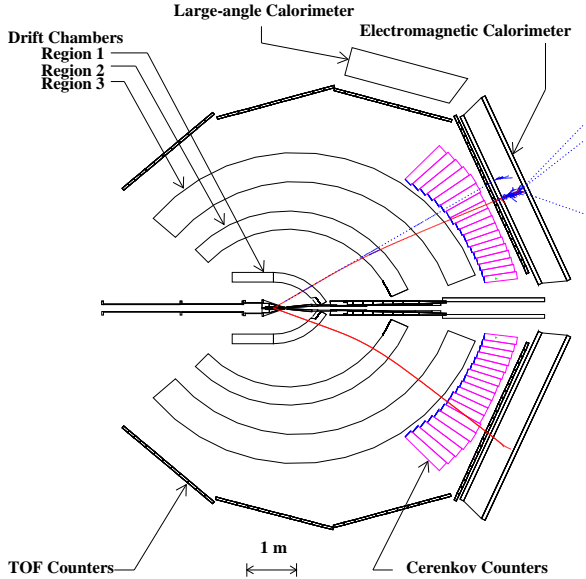}
	\includegraphics[angle=0,width=0.5\textwidth]{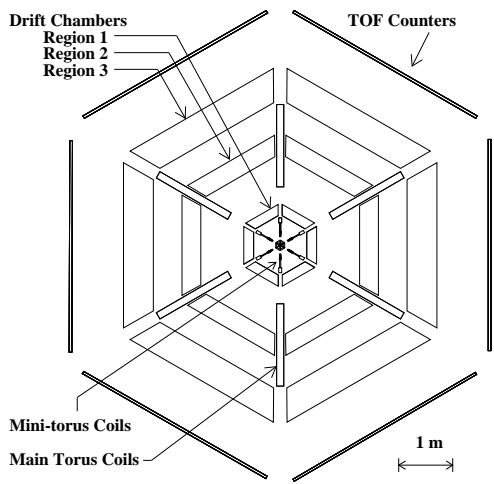}
        \caption{
	  (Color online) Schematics of the CLAS detector system. The top panel shows a horizontal cut through sectors 1 (upper hemisphere) and 4 (lower hemisphere) along the beam line. The beam enters from the left into CLAS. A GEANT-simulated event is shown with an electron bending towards the beam line and a positive particle in the opposite sector bending away from the beam. The bottom panel shows a cut perpendicular to the beam line through the center of CLAS.   
          \label{fig:clas}
        }
\end{center}
\end{figure}

The measurement was carried out with the CEBAF Large Acceptance Spectrometer (CLAS)~\cite{clas}. A schematic view of CLAS is shown in Fig.~\ref{fig:clas}. CLAS utilizes a magnetic field generated by six flat superconducting coils (main torus) in an azimuthally symmetric arrangement. The coils generate an approximately toroidal field distribution around the beam axis. The six sectors of the magnet are independently instrumented with 34 layers of drift cells for particle tracking, plastic scintillation counters for time-of-flight (TOF) measurements, gas threshold Cherenkov counters (CC) for electron/pion separation and triggering purposes, and a scintillator-lead sampling array (electromagnetic calorimeter or EC) for photon and neutron detection, as well as triggering. To aid in electron/pion separation, the EC is segmented into an inner part facing the target and an outer part further away from the target. 

 CLAS covers on average 80\% of the full $4\pi$ solid angle for the detection of charged particles in the laboratory frame. Azimuthal angle acceptance is maximum at large polar angles and decreases at forward angles. Polar angle coverage ranges from about $8^{\circ}$ to $140^{\circ}$ for the detection of $\pi^+$. Electrons are detected in the CC and EC for polar angles from $15^{\circ}$ to $55^{\circ}$, with this range being somewhat dependent on the momentum of the scattered electron and the magnetic field strength. 

 The target is surrounded by a small toroidal magnet (mini-torus) with non-superconducting coils. This magnet is used to shield the drift chambers closest to the target from the intense low-energy electron background resulting from M$\o$ller electron-scattering processes. 
In the current experiment, only two charged particles need to be detected, the scattered electron and the produced $\pi^+$, while the full final state is reconstructed using four-momentum conservation constraints. The continuous electron beam provided by CEBAF is well suited for measurements involving two or more final state particles in coincidence, leading to very small accidental coincidence contributions of  $< 10^{-3}$ for the instantaneous luminosity of $10^{34}\;\rm{cm}^{-2}\rm{sec}^{-1}$ used in this measurement.   

 The specific experimental data set used for this analysis was collected from October 2001 to January 2002, during the CLAS {e1-6} run period. The incident beam had an average intensity of $7\;\rm{nA}$ and an energy of $5.754\;\rm{GeV}$. The liquid hydrogen target was $5\;\rm{cm}$ long and located $4\;\rm{cm}$ upstream from CLAS center. The main torus magnet was set at 90\% of its maximum field. Empty-target runs were performed to measure contributions from the target cell windows. We compared our optimized beam energy of 5.754 GeV with the energy calibrated by Hall-A, which is based on concurrent high-resolution elastic electron-proton scattering measurements. Both beam energies agree within less than 6 MeV.

 Raw data were subjected to the calibration and reconstruction procedures that are part of the standard CLAS data-analysis chain. The reaction studied in this paper contributed to only a small fraction of the total event sample, and a more stringent event selection (``skimming'') was applied to select events with one electron candidate and only one positively charged track. These events were subject to further selection criteria described in the following sections. The kinematic range and bin size were optimized according to the available statistics in the covered kinematic range.  Table \ref{tab:kine_range} reports the kinematic range, bin size, and number of bins for the relevant variables. Acceptance and radiative corrections were specifically calculated for each bin given in Table \ref{tab:kine_range}. The cross section is calculated by multiplying by the radiative correction factors and dividing by the acceptance correction factors. The $\theta_{\pi}^*$ is the polar angle of the detected positive pion in the center-of-mass frame.

\begin{table}[htb]
\caption{The ranges of kinematical bins used in this analysis.
}
\begin{center}
\begin{tabular}{cccc}
\hline
Variable & Number & Range & Bin size \\
 &  of bins &  & \\
\hline
\hline
$W$  & 3  & $1.11\;-\;1.15\;\rm{GeV}$ & $20\;\rm{MeV}$ \\
$Q^2$ & 5 & $2.12\;-\;4.16\;\rm{GeV^2}$ & variable \\
$\cos\theta_{\pi}^*$ & 10 & $-1.0\;-\;+1.0$ & 0.2\\ 
$\phi_{\pi}^*$ & 12 & $0^{\circ} \sim 360^{\circ}$ & $30^{\circ}$ ($\cos\theta_{\pi}^*\ge-0.1$) \\
                 &  6 & $0^{\circ} \sim 360^{\circ}$ & $60^{\circ}$ ($\cos\theta_{\pi}^*<-0.1$)\\
\hline
\end{tabular}
\label{tab:kine_range}
\end{center}
\end{table}
%
%
%
%
%
%
%
\section{Data analysis}
\subsection{Particle identification and corrections}
\begin{figure}[!htb]
\vspace{75mm}
\centering{\includegraphics{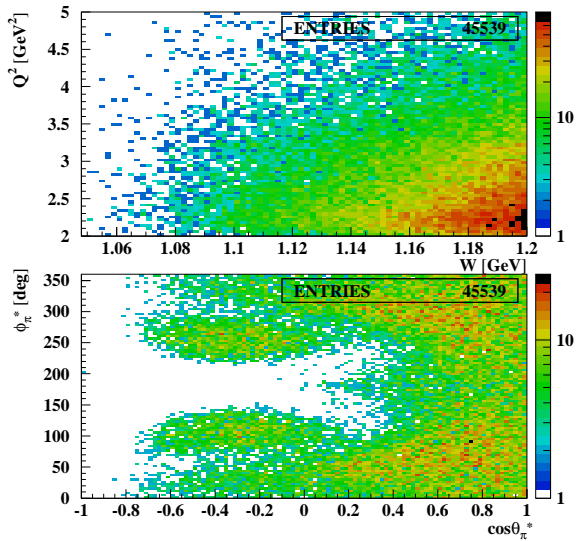}}
        \caption{
        (Color online) The kinematical coverage for pions used in this analysis in terms of the momentum transfer $Q^2$ versus the center-of-mass energy $W$ (top) and $\phi^*_{\pi}$ versus $\cos\theta_{\pi}^*$ (bottom). 
        }
        \label{fig:kinecoverage}
\end{figure}

For the particle identification (PID) and kinematic corrections we applied the standard PID cuts for the near-threshold physics regime. The total number of single-pion events with $W \le 1.2\;\rm{GeV}$ (see Fig.~\ref{fig:kinecoverage}) is approximately $4.55\times10^4$. Since PID and kinematic corrections have a strong dependence on event statistics, our PID and corrections were investigated before applying the $W \le 1.2\;\rm{GeV}$ cut to avoid  large uncertainties from such small statistics. Therefore, most of the correction procedure for electrons and pions follows the method described in the previous analysis~\cite{KPark00} with optimized parameters.
\subsubsection{Electron identification}
Electrons were tentatively identified in CLAS at the trigger level during data acquisition by requiring a minimum amount of energy deposited in the electromagnetic calorimeters (EC) in coincidence with a signal in the Cherenkov counters (CC).  This tentative identification was then improved by applying additional requirements in the off-line analysis. Coincident hits between the EC and CC were also matched with a track reconstructed in a drift chamber (DC) in the appropriate sector of CLAS. The direct correlation between the energy deposited in the EC and the momentum obtained from the track reconstruction in the DC was used to remove the residual pion contamination.

\begin{figure}[!htb]
\vspace{50mm}
\centering{
\includegraphics{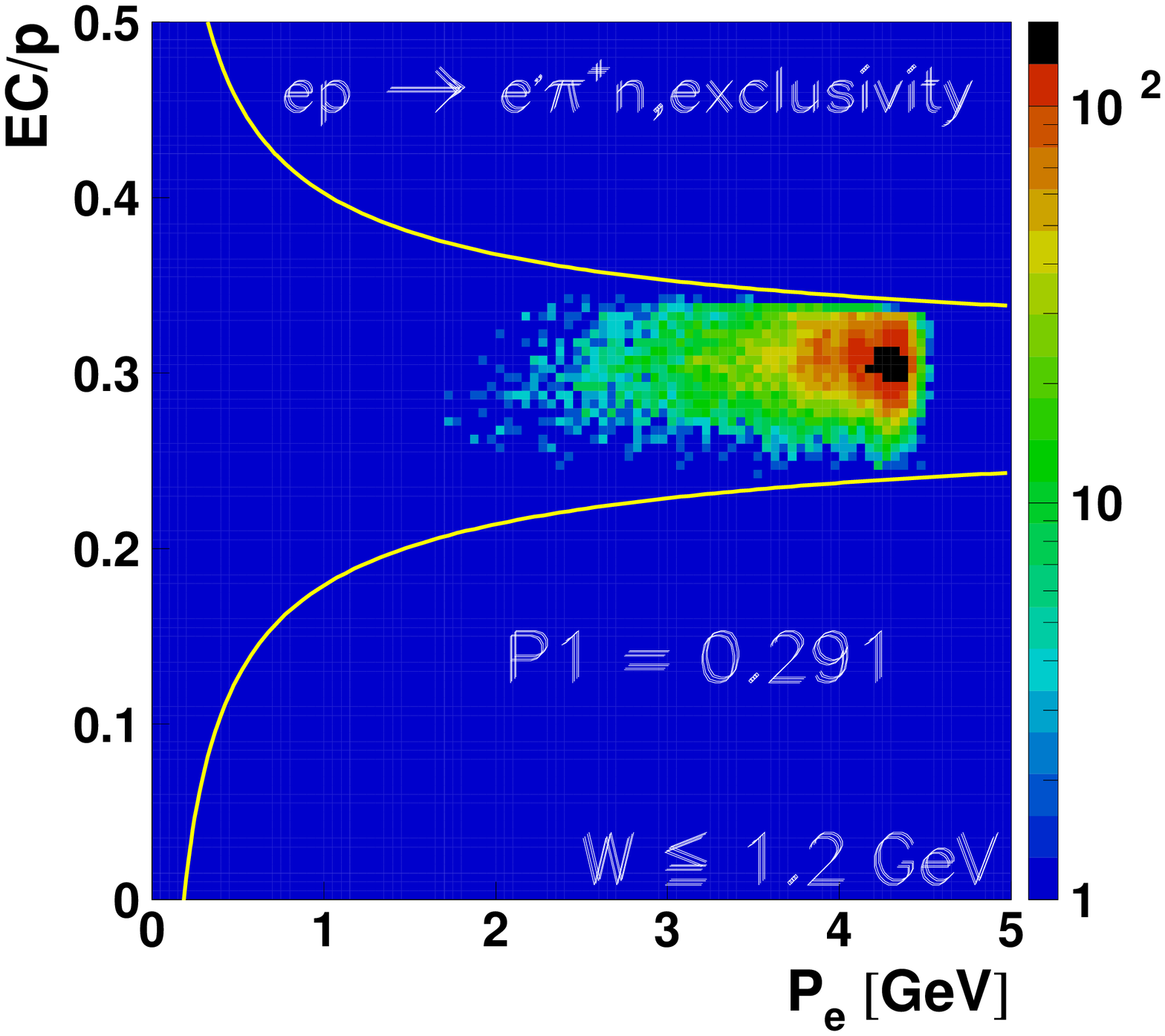}
\includegraphics{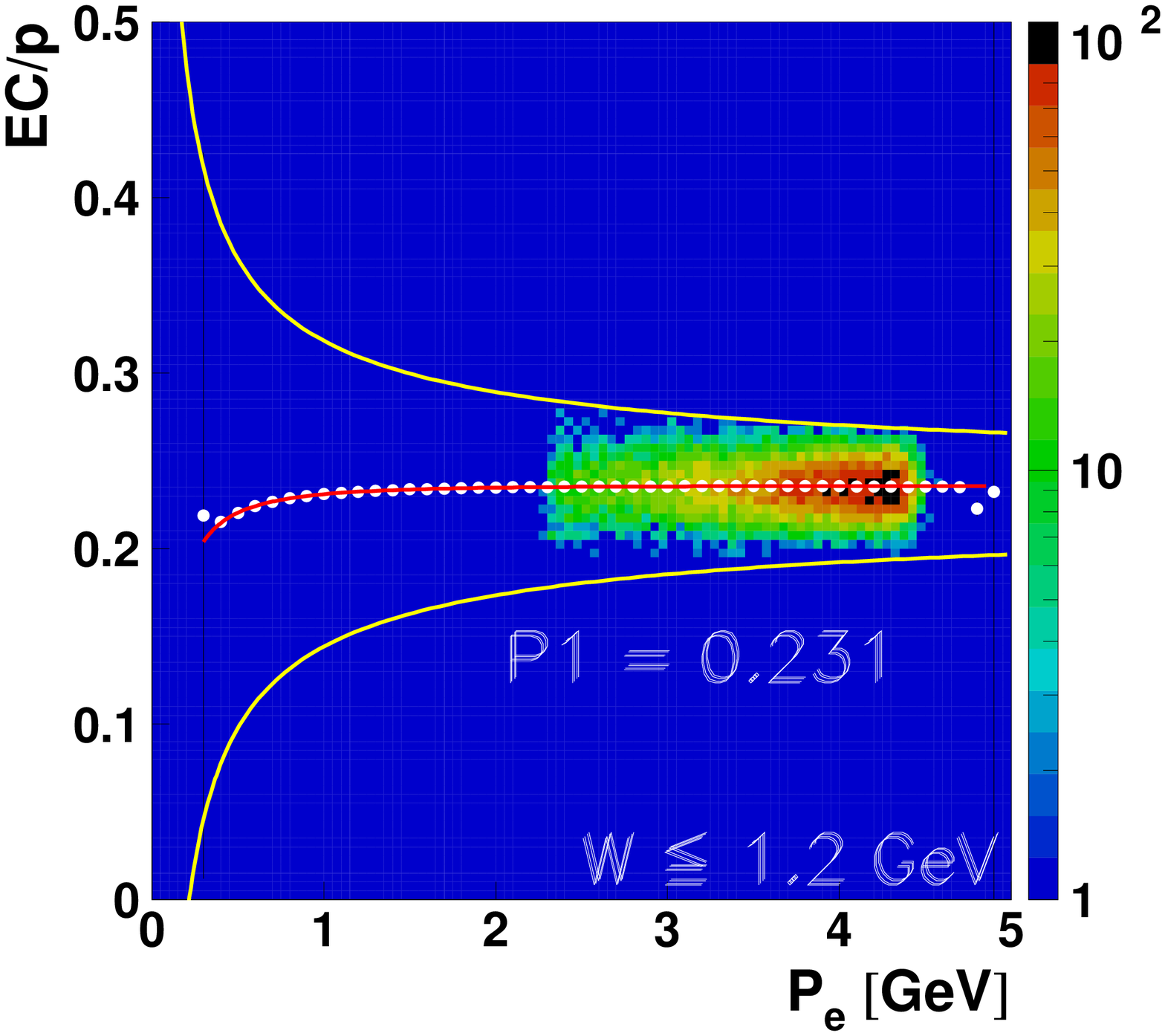}}
        \caption{
        (Color online) Optimized sampling fraction constant of EC versus electron momentum with $W \le 1.2\;\rm{GeV}$ for the experimental data (left) and Monte Carlo simulation (right). The solid lines (left) show the $\pm 3\sigma$ cuts from all electron fit. $P1$ is the sampling fraction from fit.
        }
        \label{fig:ec_momentum1}
\end{figure}

 About 30\% of the total energy deposited in the EC is directly measured in the active scintillator material; the remainder of the energy is deposited mostly in the lead sheets interleaved between the scintillator sheets as showering materials. To improve the separation of electrons and pions, the ratio E$_{tot}/p$ was used, where E$_{tot}$ and $p$ are the total energy deposited in the calorimeter and the deduced momentum for the particle, respectively. This ratio, which is called the sampling fraction ($\alpha$), is nearly momentum independent for the range of electron momenta (2.5 to 4 GeV) in this analysis. The sampling fraction for electrons (determined by all electrons over the full $W$ range) was found to be $0.291$ for this experiment, a value roughly comparable with the estimate of $0.231$ for that ratio determined in a Monte Carlo simulation. Figure~\ref{fig:ec_momentum1} shows the application of the sampling fraction cut to experimental (left) and simulated (right) data. The solid lines represent the $\pm 3\sigma$ sampling fraction cuts for the experimental data and the Monte Carlo simulation in this analysis. 

 Most of the produced pions that passed through the EC were minimum ionizing particles. These pions deposited energy both in the inner and the outer stacks of the calorimeter in amounts almost independent of their kinetic energy and only related to the detector thickness. Pions were identified by the precise correlation between the minimum total energy deposited in the calorimeter E$_{tot}$ and the energy deposited in the inner part of the calorimeter E$_{in}$. To avoid continuous triggering on noise, a minimum signal threshold was set for the calorimeter. For this experiment, the threshold was set at a level of $172$ mV, such that only electrons with momenta greater than about $640$ MeV were detected.

 A GEANT simulation (GSIM) was used to determine the response of the electromagnetic calorimeter as a function of electron energy. When an electron hit is close to the calorimeter edges, part of the shower leaks outside the device; in this case, the energy cannot be fully reconstructed from the calorimeter information alone. This problem was avoided by selecting only those electrons lying inside a fiducial volume within the electromagnetic calorimeter that excludes the detector edges.

\begin{figure}[htb]
\begin{center}
	\includegraphics[angle=0,width=0.48\textwidth]{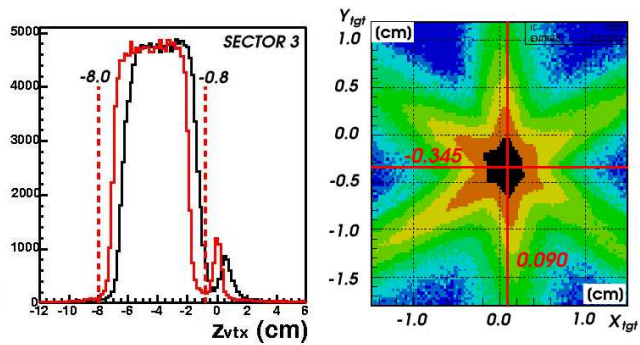}
        \caption{
          (Color online) $z$-vertex distribution for sector 3 before (black solid line) and after (red solid line) vertex correction (left). The vertex cuts are illustated by the vertical dashed lines. Reconstructed $X$ and $Y$ target positions, showing an offset of $X_{\rm{tgt}}=0.90\;\rm{mm}$ and $Y_{\rm{tgt}}=-3.45\;\rm{mm}$, respectively (right).
          \label{fig:xyzvertex0}
        }
\end{center}
\end{figure}

 Tracking information from the drift chambers was used to reconstruct for each event an originating vertex location in the target region. Particle identification was improved by eliminating events from the analysis that had reconstructed vertex positions outside the known volume of the target (which included a small target misalignment from the beam axis). For this experiment, these vertex requirements demanded that the reconstructed $z$-vertex position (distance along the beam axis from the center of CLAS, with positive values indicating downstream of the center) lie in $-80\;\rm{mm}< Z_{vtx} < -8\;\rm{mm}$. We corrected the $x$-, $y$-vertex positions for the beam centering on the target, and the $z$-vertex cut was imposed on the reconstructed vertex locations based on the beam axis. Since the beam position was not precisely centered on the target, with offsets of $X_{\rm{tgt}}=0.90\;\rm{mm}$ and $Y_{\rm{tgt}}=-3.45\;\rm{mm}$, the $z$-vertex was corrected for this small misalignment of the beam position before the $z$-vertex cut was imposed on the reconstructed vertex location.  Figure~\ref{fig:xyzvertex0} (left) shows the $z$-vertex distribution for sector 3 before and after the vertex correction, and the $z$-vetex cuts that have been applied. Figure~\ref{fig:xyzvertex0} (right) shows the transverse beam position. We also take an empty target contribution into account.

 Coincident hits between the EC and CC were also matched with a track characteristic for a negative particle that is reconstructed in the drift chambers of the same CLAS sector. A lower threshold on the number of photoelectrons detected in the photomultiplier tubes of the CC for an event provided an additional cut for improving electron identification. The number of photoelectrons detected in the CC sectors follows a Poisson distribution, modified for irregularities in light collection efficiency for the individual elements of the array. For this experiment, a good electron event was required to have more than 2.5 photoelectrons detected in CC.

\subsubsection{Pion identification}
For $\pi^+$ identification a track characteristic for positive particles has to match with a corresponding hit in the TOF system.  Pions are then separated from other positively charged particles based on their hadron velocity $\beta_h = v/c$, which is obtained from the difference between the vertex start time and the time of flight of the TOF counters (SC), and their hadron momentum $p_h$, which is determined by tracking the hadron through  the magnetic field in the drift chambers. To isolate pions from protons, a $\pm 2\sigma$ cut on $\beta_h$ versus $p_h$ is applied.

\begin{figure}[!htb]
\vspace{115mm}
\centering{
\includegraphics{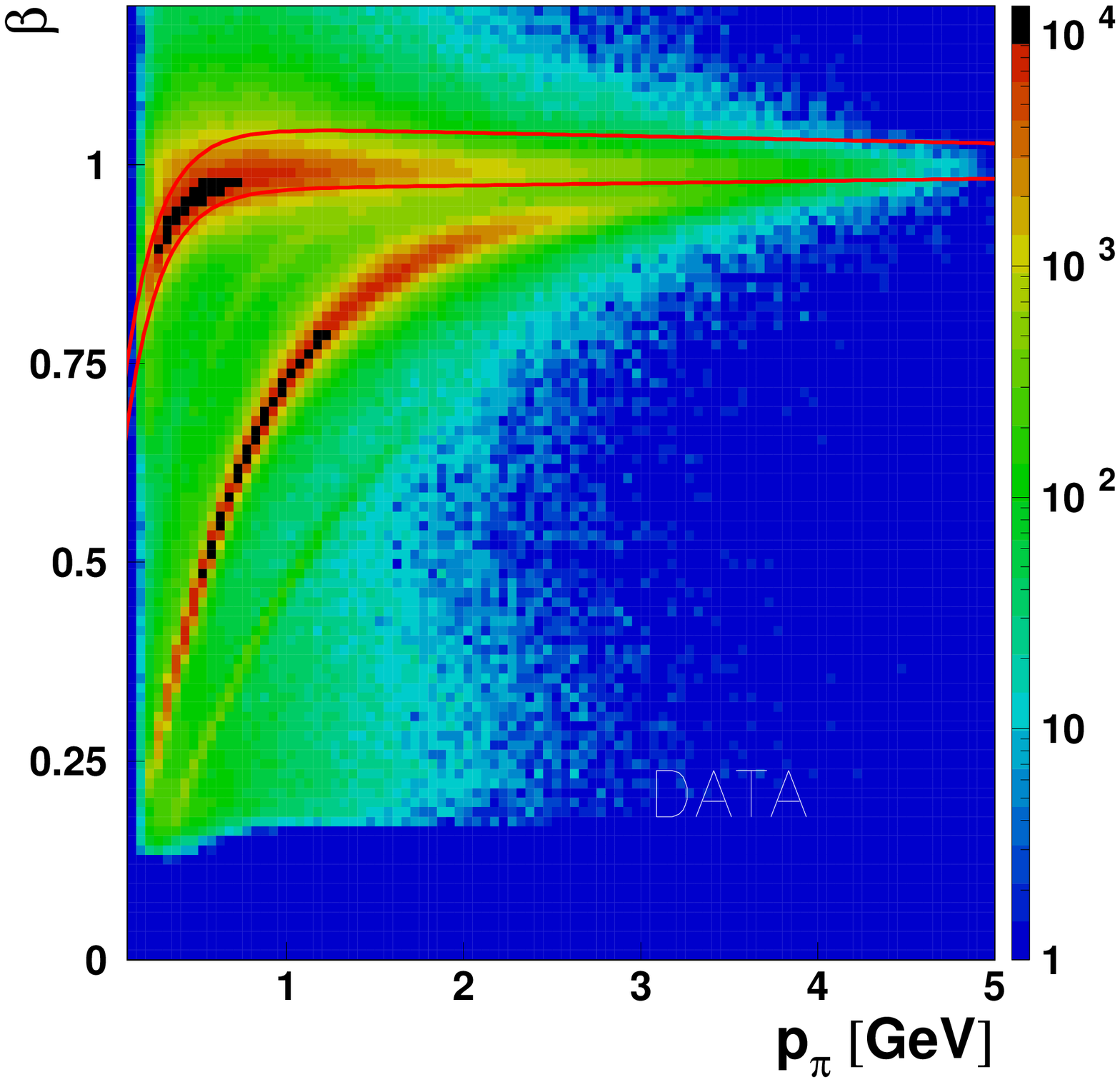}
\includegraphics{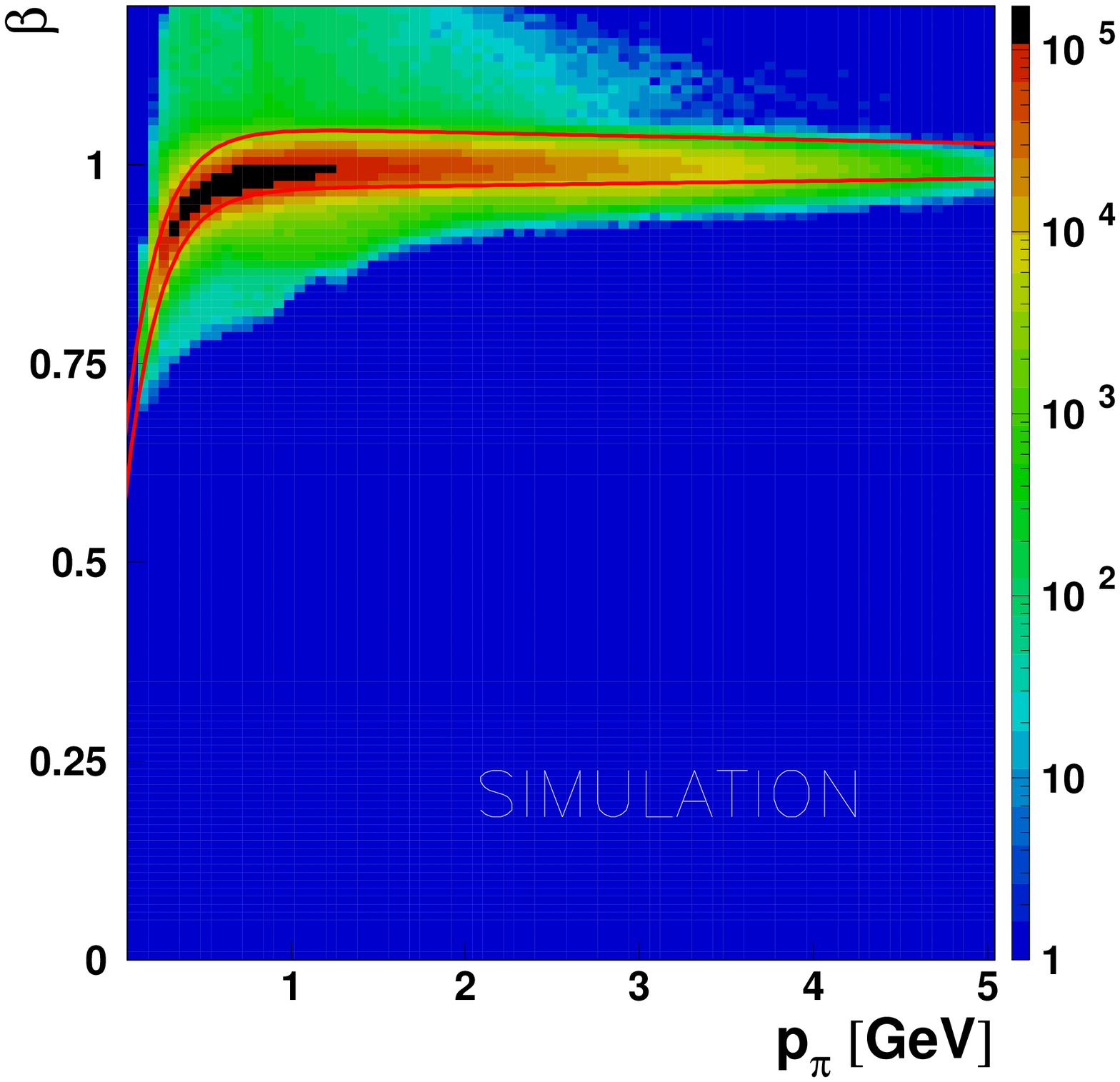}}
        \caption{
	 (Color online)  The particle velocity distribution $\beta$ versus particle momentum. 
The pion velocity sample is fit by a function of the form $A+\exp(B+ C p_{\pi})$ to generate the $\beta$ cut for both simulation (upper) and experimental data (lower).
        }
        \label{fig:pion_id_beta_mean}
\end{figure}

Figure~\ref{fig:pion_id_beta_mean} shows the particle's velocity versus  momentum for positive tracks. A single Gaussian fit function was applied to $\beta_h$ in each momentum bin to choose the proper velocity cut for pion identification. The solid red lines superimposed on the scatter-plots in Fig.~\ref{fig:pion_id_beta_mean} show the $\beta_h$ cuts for both simulation (upper) and data (lower). All remaining positrons were considered as pions due to the limited momentum resolution, which increases the background. However, the missing mass and vertex cuts reduce the background to a few percent.

We observed that the other positive charged particles (protons, kaons) contamination under the missing mass peak is negligibly small, even at high pion momenta. Overall, background contributions under the neutron missing mass are negligible in our kinematic region; hence, no background subtraction was performed.

\subsection{Fiducial cuts}
 Due to the complexity of the CLAS geometry and edge effects, we define a fiducial volume that is restricted to detector regions with nearly full particle acceptance and high reconstruction efficiency. We made individual CLAS sector dependent geometrical fiducial cuts and applied the same cuts to the simulations and to the experimental data, where those cuts select areas of uniform detector response that can be reproduced by Geant3 based simulation (GSIM) with CLAS detetor geometric information. In order to implement the sector fiducial cuts, the GSIM program only requires knowledge of the momentum and charge of the particle, since the curvature of the trajectory of the particle depends only on the particle's charge and momentum, and the strength of the CLAS magnetic field. The fiducial cuts have been applied to both electrons and pions in the same way as in the previous analysis~\cite{KPark00} using optimized parameters.

\subsubsection{Electron sector fiducial cuts}
\begin{figure}[!htb]
\vspace{60mm}
\centering{
\includegraphics{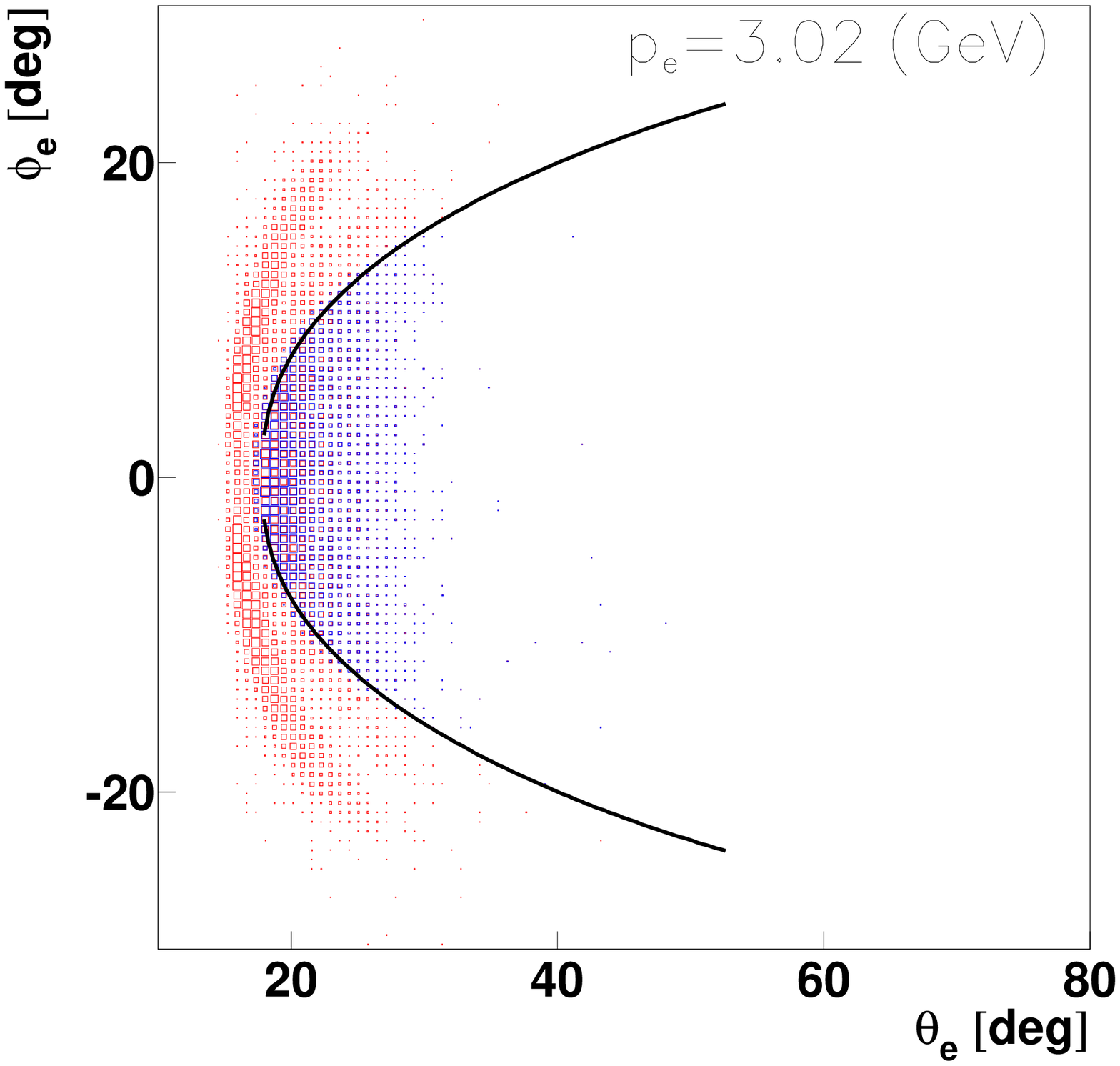}
\includegraphics{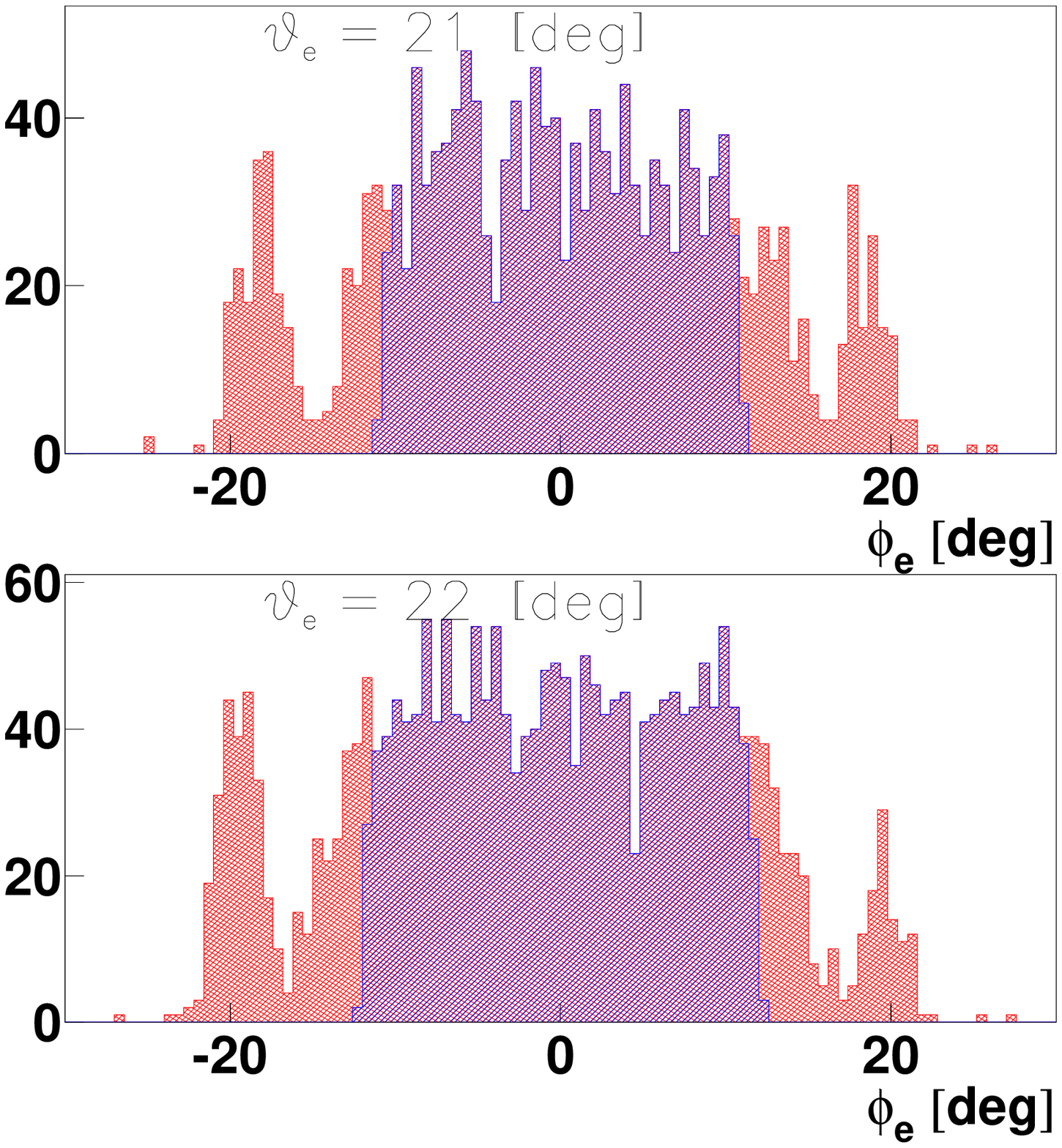}}
        \caption{
	 (Color online) The electron angular distribution in a sample momentum bin $p_e=3.02\pm0.125\;\rm{GeV}$ for sector 4 (left). The black solid curve indicates the fiducial cut boundary.  The criterion to determine the electron fiducial region in terms of $\phi_e$ for one momentum bin is the detector efficiency. In order to eliminate the depleted regions of the detector we chose the flat area by looking through the $\theta$-sliced $\phi_e$ distributions. The right plots show examples of the $\phi_e$ distribution in two $\theta_e$ bins: $21^{\circ}$ and $22^{\circ}$ for sector 4. The highlighted areas around the center indicate the selected fiducial range.
        }
        \label{fig:e_fcut_loose00}
\end{figure}

 Since the toroidal magnetic field bends the electrons inward, the fiducial cut in polar angle $\theta_e$ and azimuthal angle $\phi_e$ depends on the momentum of the electron $p_e$~\cite{KPark00}. The azimuthal symmetry of the angular distribution for the electrons was considered when selecting these regions. Therefore, for fixed $\theta_e$ and $p_e$, one expects to find a flat distribution in $\phi_e$. Several cut functions to eliminate depleted regions in momentum $p_e$ and angle $\theta_e$ were applied empirically. Figure~\ref{fig:e_fcut_loose00} shows such a sector centroidal electron angle $\phi_e$ distribution in a sample momentum bin of $3.02\pm0.125\;\rm{GeV}$. The black solid curve in the left plot shows the boundary of the fiducial region for the central momentum in that bin. Only events with electrons inside the black curve are used in the analysis. In addition, a set of $\theta_e$ versus $p_e$ cuts are used to eliminate the areas with a depleted number of events due to bad time-of-flight counters, Cherenkov counter photomultiplier tubes, or drift chamber wires.

\subsubsection{Pion sector fiducial cuts}\label{pion_fidu}
 Pion fiducial cuts are designed to exclude regions of non-uniform acceptance due to interactions with the mini-torus coils, torus cryostat, or the edges of the drift chambers. The onset of these regions is not symmetric about the sector mid-plane but is sector as well as momentum dependent. This cut is a function of the pion momentum $p_{\pi}$ after selecting good electrons. 

\begin{figure}[!htb]
\begin{center}
        \includegraphics[angle=0,width=0.38\textwidth]{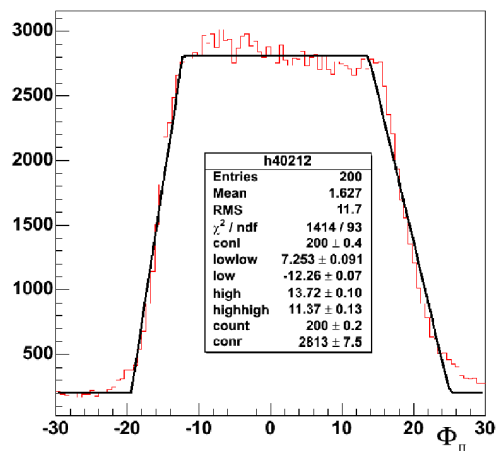}
        \caption{
          An example of a $\phi_{\pi}$ (deg) fit with a trapezoidal function. The flat region was used to select the sector fiducial cut for
the kinematical bin with $0.35\;\rm{GeV} <  p_{\pi} <0.45\;\rm{GeV}$ and $23^o <  \theta_{\pi} < 25^o$ bin for sector 4. 
          \label{fig:parameter_fidu}
        }
\end{center}
\end{figure}

 The pion momentum was scanned with a $100\;\rm{MeV}$ bin width from $0.3\;\rm{GeV}$ to $1.7\;\rm{GeV}$. The pion angular distribution was investigated in each momentum bin. The $\pi^+$ fiducial cut functions are parametrized by the pion momentum and the angles $\theta_{\pi}$ and $\phi_{\pi}$. The same pion cuts are applied to data as well as simulation.  To define the reasonable detector response region, the two-dimensional plot of $\phi_{\pi}$ versus $\theta_{\pi}$ was sliced along  $\theta_{\pi}$ in $2^{\circ}$ bins for $12^{\circ} \le \theta_{\pi} \le 100^{\circ}$, and each $\phi_{\pi}$ distribution was fit by a function which included a trapezoidal shape and a constant. A fit example is shown in Fig.~\ref{fig:parameter_fidu}. Each momentum and $\theta_{\pi}$ bin in each sector has a unique $\phi_{\pi}$ plateau region. The corresponding fit parameters are functions of sector, $p_{\pi}$, and $\theta_{\pi}$. The correlation between $\phi_{\pi}$ and $\theta_{\pi}$ is described by an exponential and a third order polynomial function.

\subsection{Kinematic corrections}\label{Kinematics_Corrections}
\begin{figure}[htb]
\begin{center}
	\includegraphics[angle=0,width=0.42\textwidth]{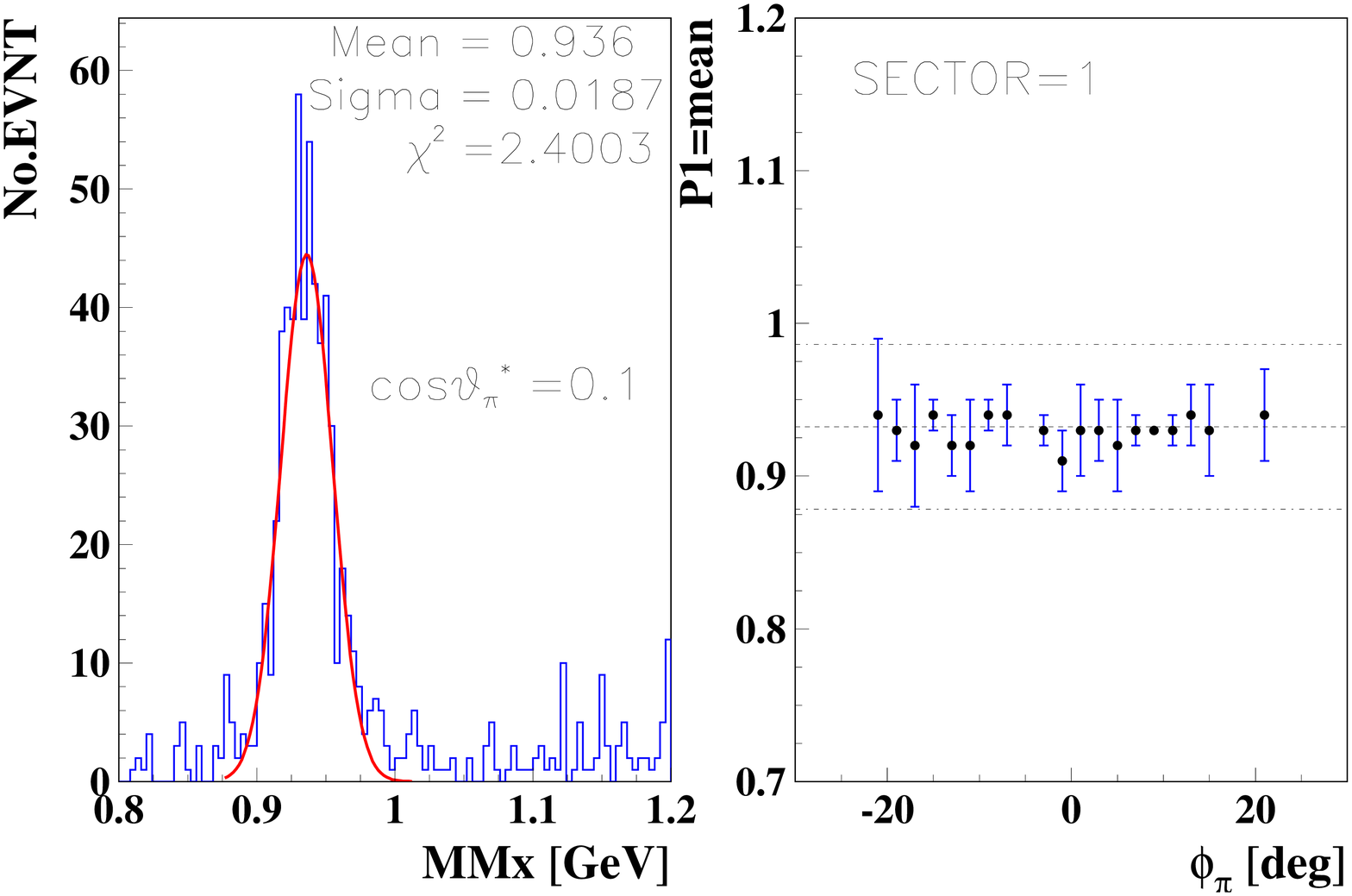}
        \caption{
          An example of a neutron missing mass fit for sector 1 by a single Gaussian function in the bin at $\cos\theta_\pi=0.1$, $W =1.11\;\rm{GeV}$ and $Q^2=2.44\;\rm{GeV^2}$ (left), and the neutron missing mass $\phi_\pi$ dependence for sector 1 (right). Dashed lines are defined in the text. 
          \label{fig:kinecorr1}
        }
\end{center}
\end{figure}

 The kinematic corrections from the previous analysis~\cite{KPark00} were applied in this analysis. The corrections are based on the measured angles and momenta of the detected particles (electrons and pions). Figure~\ref{fig:kinecorr1} (left) shows the neutron missing mass for $W=1.11\;\rm{GeV}$, which is the lowest $W$ bin that we can access in this data set. Overall, there is no systematic $\phi_{\pi}$-dependence on the inferred neutron mass. The central horizontal dashed line in the right-hand plot of Fig.~\ref{fig:kinecorr1} indicates the mean value found for the neutron mass in the left-hand plot; the upper and lower horizontal dashed lines indicate $\pm3\sigma$ of the missing mass distribution, where $\sigma$ is the energy resolution as defined by the standard deviation.
\vspace{5mm}

\section{Monte Carlo simulations}\label{Monte_Carlo_Simulation_Data}
  The detection efficiencies for various particles in CLAS were estimated from Monte Carlo simulations of the detector. We used the two different event generators, AAO-RAD (a physics-model-based event generator~\cite{maid2003}) and GENEV (a phase space event generator~\cite{GENEV_GEN}). Both event generators generate exclusive events including radiative effects. The AAO-RAD event generator uses the MAID2003 model~\cite{maid2003}. The GENEV event generator allows us to generate several exclusive electroproduction reactions, from pion production to the production of vector mesons ($\omega$, $\rho^0$, and $\phi$) including their decays, as well as non-resonant multi-pion production. Cross section tables for these processes are used that are based on photoproduction data and extrapolated to the case of electroproduction.

  Several million (AAO-RAD) and $\sim 50$ million (GENEV) single-pion events were generated near the pion threshold region between $1.1\;\rm{GeV} \le$$W$$\le 1.2\;\rm{GeV}$ and $1.0\;\rm{GeV}^2 \le$$Q^2$$\le 10.0\;\rm{GeV}^2$ with full angular coverage. Both generated data sets were processed by the standard GEANT simulation of the CLAS detector. The statistical uncertainties for both simulations are less than a percent. Most steps closely followed the previous analysis described in Ref.~\cite{KPark00}.

\begin{figure}[!htb]
\vspace{100mm}
\centering{\includegraphics{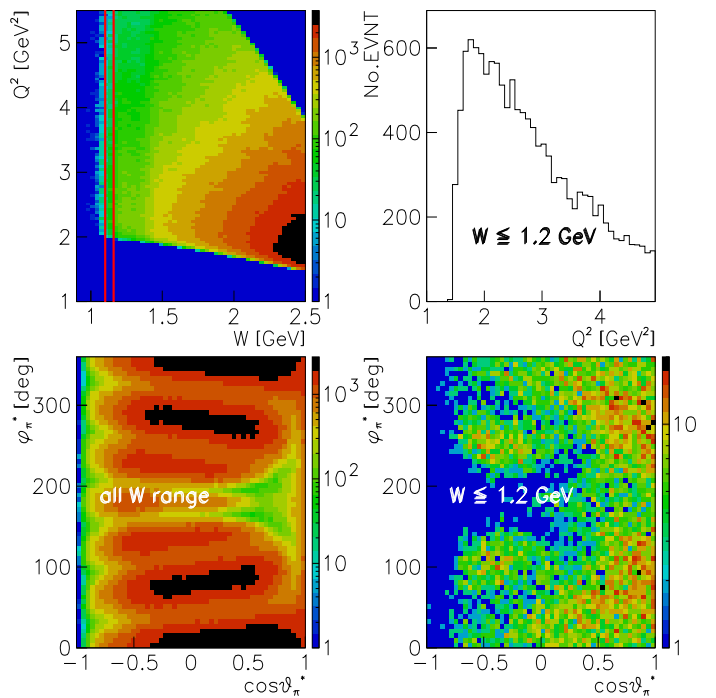}}
        \caption{
          (Color online) Examples of Monte Carlo simulation results: $Q^2$ versus $W$ and the angular distribution $\phi^*_{\pi}$ versus $\cos\theta_{\pi}^*$ for all $W$ (left column), and yield versus $Q^2$ and the angles $\phi^*_{\pi}$ versus $\cos\theta_{\pi}^*$ for $W\le1.2\;\rm{GeV}$ (right column) as a snap shot after GSIM reconstruction but still without the GPP step. The two vertical solid lines on the top-left plot show the $W$ region between $1.10\;\rm{GeV}$ and $1.16\;\rm{GeV}$ of interest.
          \label{fig:gsim_snap}
        }
\end{figure}

 The comparison of the detection efficiency results using the two different event generators indicates a $4\%$ uncertainty in the simulated efficiency prediction. Scattered electron events were simulated by AAO-RAD and GENEV for exclusive single charged pion electroproduction on a liquid hydrogen target in CLAS for an incident electron energy of $5.754\;\rm{GeV}$ and a CLAS torus current of 3375 A. Figure~\ref{fig:gsim_snap} shows examples of results from the simulations of $W$, $Q^2$, and angular distributions. 

%
The GSIM Post-processor (GPP) is used for fine adjustments of the reconstructed GSIM data to better match the measured data. GSIM simulates events in an ideal detector system and GPP is used to adjust two quantities. One is the drift chamber position resolution smearing factors that affect the tracking momentum resolution. The other is the TOF time smearing factor that affects the timing resolution. The optimized GPP settings were taken from Ref.~\cite{KPark00}. 


 The GPP output has been processed with the event reconstruction software. In order to relate experimental yields to  cross sections, one needs to calculate the acceptance, including the efficiency of the detector and radiative effects. Both acceptance and radiative corrections are processed similarly as in the previous analysis~\cite{KPark00}.

\subsection{Acceptance correction}\label{Acceptance_Corrections}
\begin{figure}[!htb]
\vspace{45mm}
\centering{\includegraphics{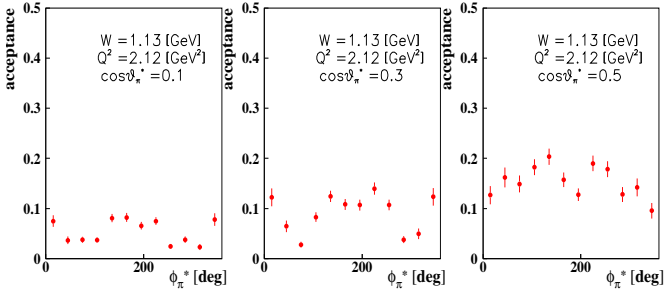}}
        \caption{
         Examples of the $\phi^*_{\pi}$-dependent acceptances for the $\gamma^*p \to n\pi^+$ reaction at $W=1.13\;\rm{GeV}$, $Q^2=2.12\;\rm{GeV^2}$, and $\cos\theta_{\pi}^*=0.1 - 0.5$. 
        }
        \label{fig:acc00}
\end{figure}

To calculate the acceptance for the $\pi^+n$ channel in the CLAS detector system, one can define the acceptance in a given kinematic bin $i$ as
\begin{eqnarray}\label{eq:ratio2}
\mbox{Acceptance}_i = \frac{N_i^{REC}}{N_i^{GEN}}~,\nonumber
\end{eqnarray}
where $N_i^{GEN}$ is the  number of events generated and $N_i^{REC}$ is the number of events reconstructed after imposing all cuts. Figure~\ref{fig:acc00} shows typical examples of $\phi^*_{\pi}$-dependent acceptances using the AAO-RAD event generator.

\subsection{Radiative corrections}\label{Radiative_Corrections}
 The radiative corrections (RC) were calculated using ExcluRad~\cite{afanasev}. ExcluRad gives an estimate calculation of radiative effects in given kinematic bins. The general definition of the radiative correction factor is the ratio of radiative events to events without radiative effects in a given kinematic bin. It is defined in a fixed $W$, $Q^2$, $\theta_{\pi}^*$, and  $\phi^*_{\pi}$ bin ($j$): $RC_j = \big({N_j^{RAD}}/{\int \sigma^{RAD}}\big)\cdot\big({\int \sigma^{NORAD}}/{N_j^{NORAD}}\big)$, where $RC_j$ is the radiative correction for bin ($j$), $\int \sigma^{RAD}$ is the radiated model cross section and $\int \sigma^{NORAD}$ the un-radiated model cross section with integrated luminosity. $N_j^{RAD}$ and $N_j^{NORAD}$ are event numbers for radiative and non-radiative events, respectively. The detailed procedure of radiative corrections is described in ~\cite{KPark00}. 

\begin{figure}[htb]
\begin{center}
	\includegraphics[angle=0,width=0.50\textwidth]{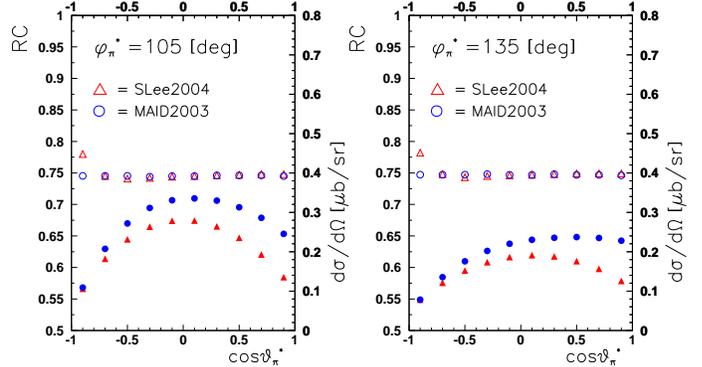}
        \caption{
          (Color online) Cross sections (solid symbols) and radiative correction factors (open symbols) as functions of $\cos\theta_\pi^\ast$ in a fixed bin centered at $W=1.15\;\rm{GeV}$ and $Q^2=2.91\;\rm{GeV^2}$. The two plots are for different values of the azimuth, $\phi^*_{\pi} =105^\circ$ and $135^\circ$. Each plot shows the RC factor on the left $y$-axis and differential cross sections on the right $y$-axis. The (red) triangles (open and closed) are from the Sato-Lee2004 model~\cite{satolee02}. The (blue) circles (open and closed) are from the MAID2003 model.
          \label{fig:frad00}
        }
\end{center}
\end{figure}

 Since there are several models to describe the cross sections near the threshold, it is important to verify whether ExcluRad gives consistent correction factors independent of the physics models. In order to perform this study, we used two different models, which are Sato-Lee2004~\cite{satolee02} (dynamic model) and MAID2003~\cite{maid2003} (unitary isobar model) as ExcluRad inputs. Both cover the kinematic region of interest. Figure~\ref{fig:frad00} shows examples of the radiative-correction comparison between the two models, as a function of $\cos\theta_{\pi}^*$ at a fixed $W=1.15\;\rm{GeV}$ and $Q^2=2.91\;\rm{GeV^2}$ but for different $\phi^*_{\pi}=105^{\circ}, 135^{\circ}$. As Fig.~\ref{fig:frad00} shows, the radiative corrections from both models are consistent even where their cross sections are quite different.

\section{Results}
\subsection{Differential cross sections  }
In order to extract the differential cross sections, all previously described corrections, efficiencies, cuts, photon flux and luminosity normalizations have to be applied to the data.  The azimuthal angle $\phi^*_{\pi}$ is divided into $60^\circ$ bins for $\cos\theta_\pi^\ast <-0.1$ (yielding 6 bins) and $30^\circ$ bins for $\cos\theta_\pi^\ast\ge -0.1$ (yielding 12 bins). This binning and a minimum acceptance cut ($> 0.6\%$) limit the statistical uncertainties.

\begin{figure}[!htb]
\vspace{100mm}
\centering{
\includegraphics{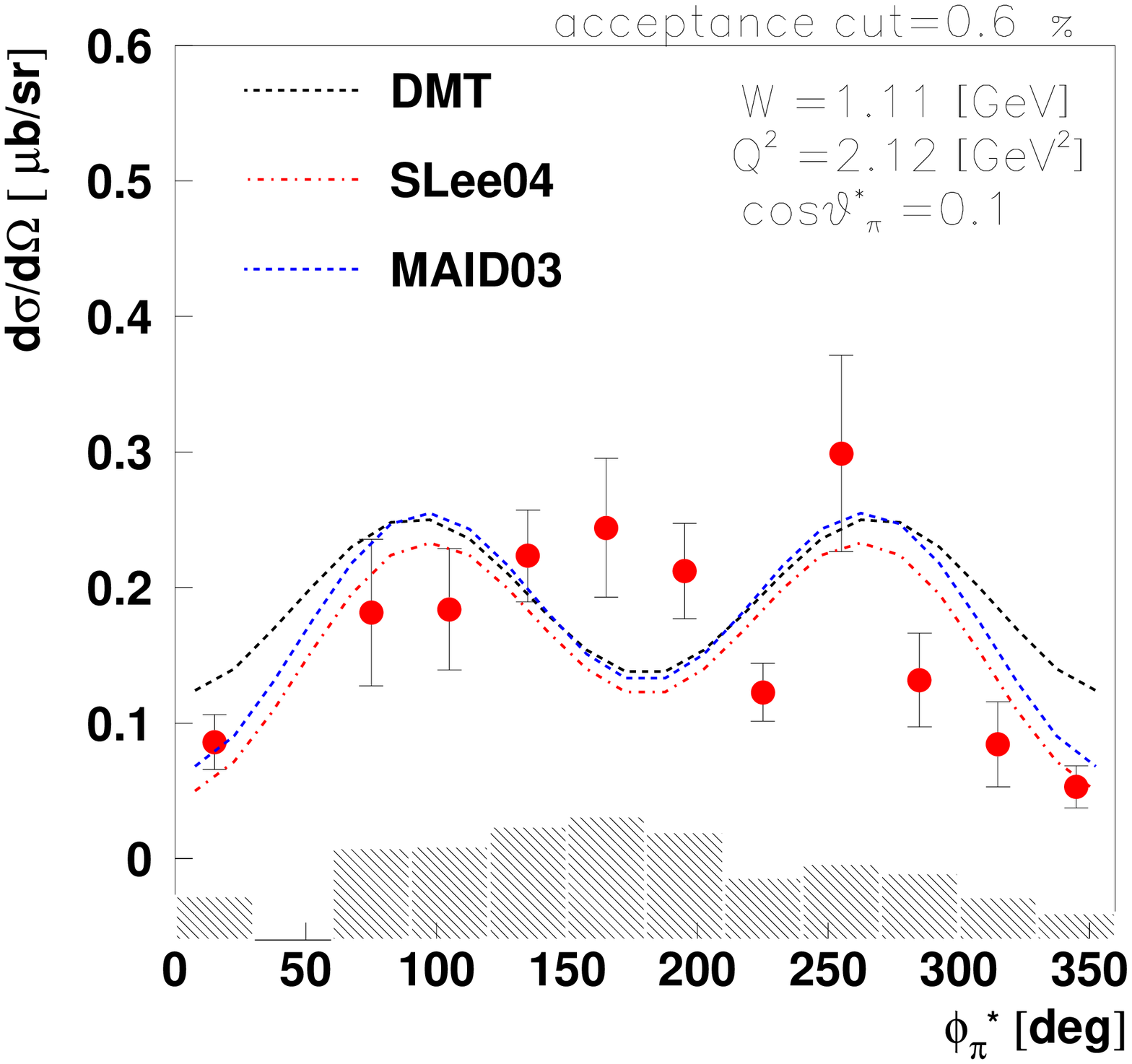}
\includegraphics{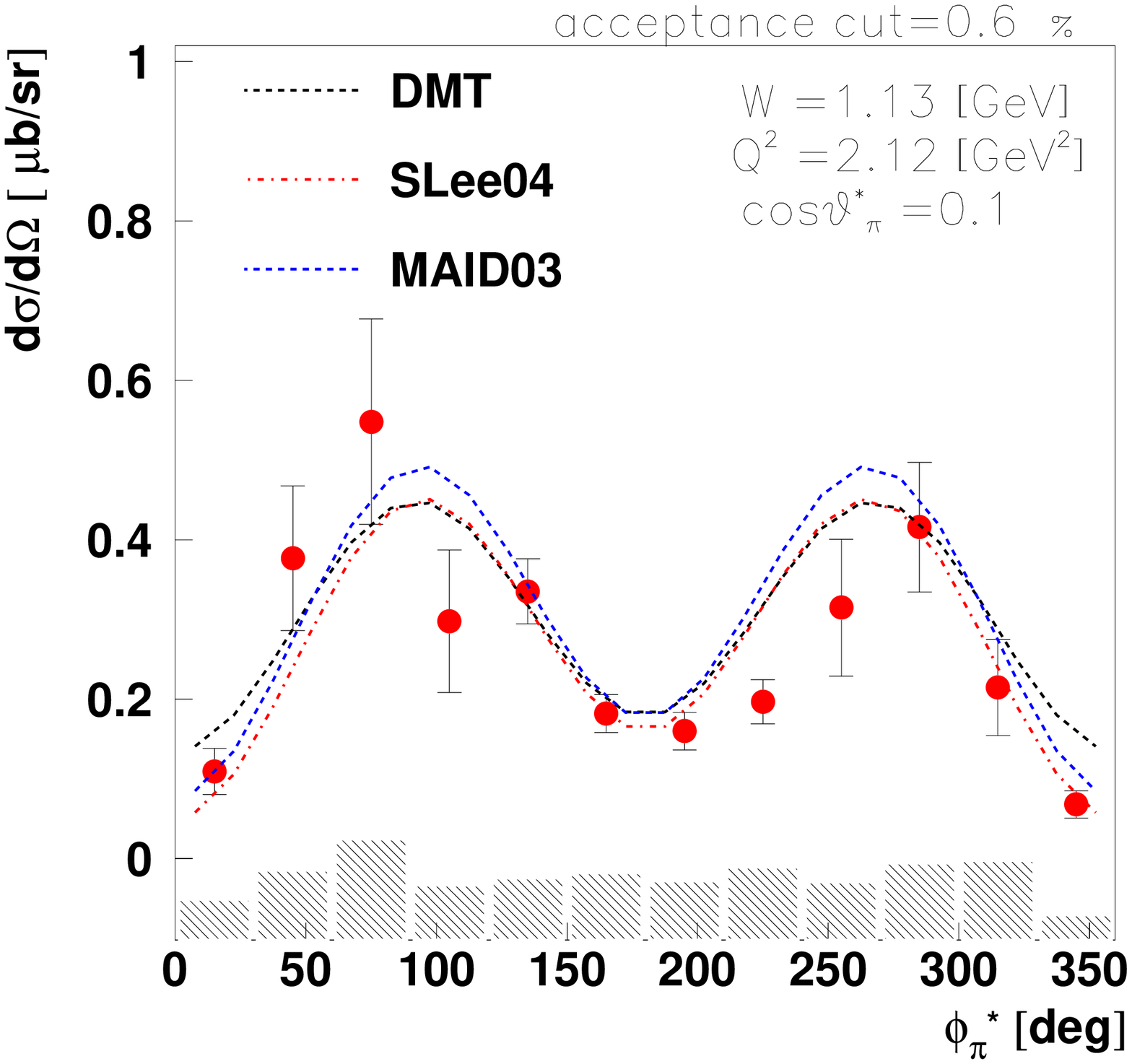}
\includegraphics{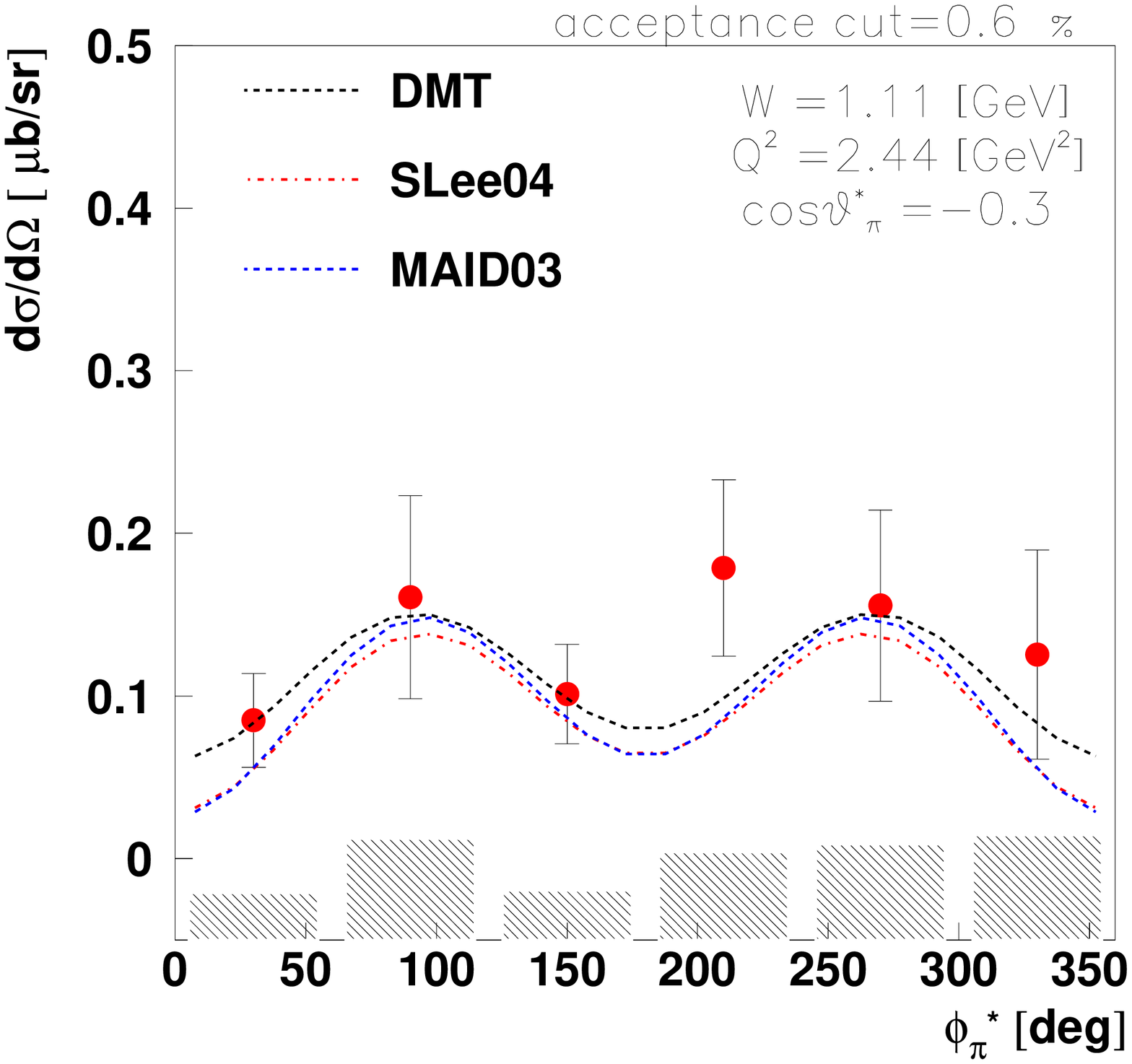}
\includegraphics{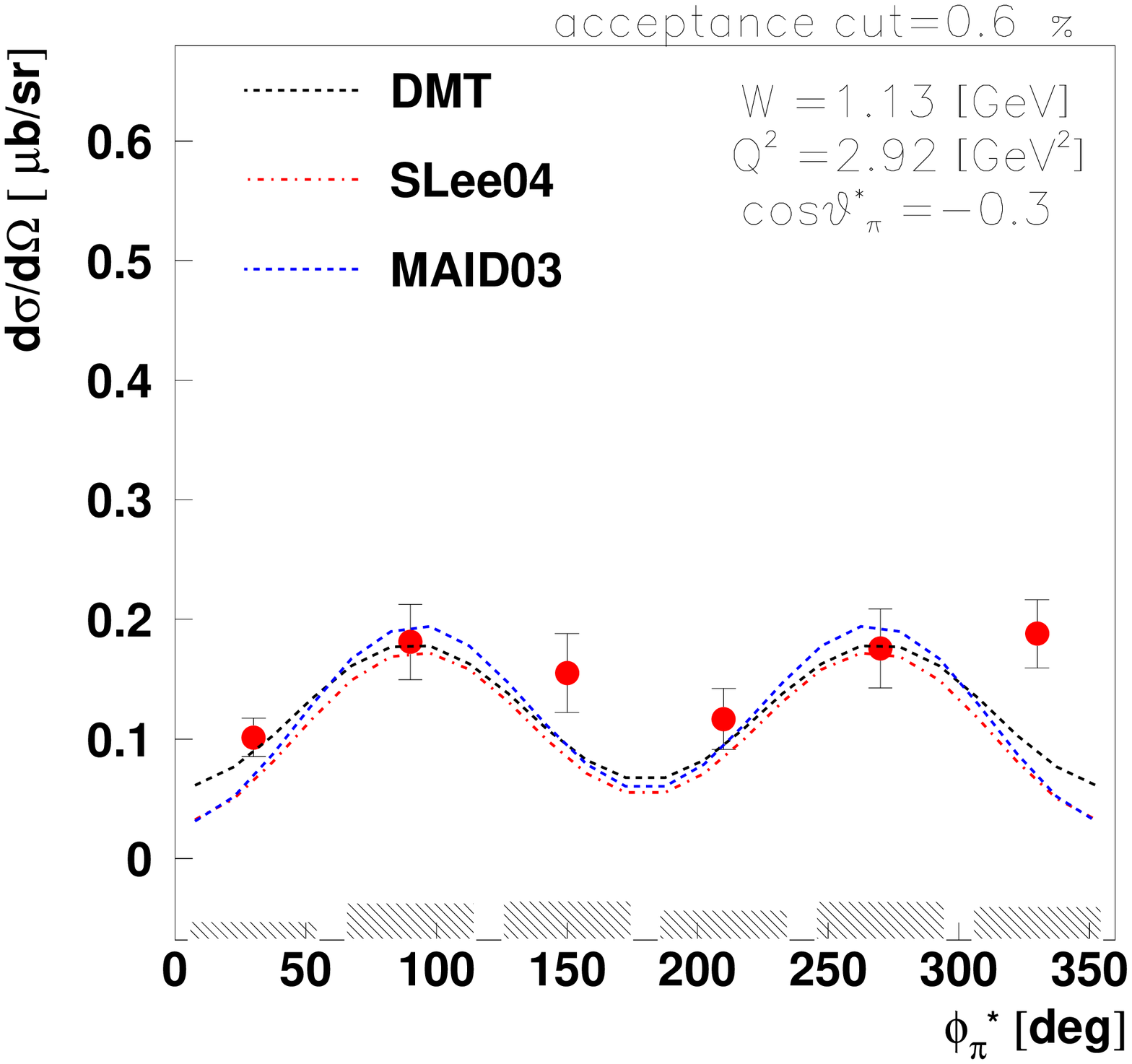}}

        \caption{
        (Color online) Examples of the differential cross sections as function of $\phi^*_{\pi}$ with $0.6\%$ minimum acceptance cut at $W=1.11$, $1.13\;\rm{GeV}$, $Q^2=2.12$, $2.44$, $2.92\;\rm{GeV^2}$, and $\cos\theta_{\pi}^*=0.1$, $-0.3$. The error bars of the data include only statistical errors and the shaded bars show the systematic uncertainties. Results from the DMT model~\cite{dmt00} are indicated by the black line. Also shown are the Sato-Lee 2004 model (red dashed-dot line)~\cite{satolee02} and MAID2003 (blue dashed line)~\cite{maid2003}.  
        }
        \label{fig:crs01}
\end{figure}

 Figure~\ref{fig:crs01} shows an example of the $\phi^*_{\pi}$-dependent differential cross section (left) and the corresponding fit (right) in a $W=1.11\;\rm{GeV}$ bin that is closest to the pion threshold in this analysis, with comparisons to the Dubna-Mainz-Taipai (DMT)~\cite{dmt00}, Sato-Lee 2004~\cite{satolee02}, and MAID2003~\cite{maid2003} models. The $\chi^2$ is calculated by ${\chi_0^2}/{(N_{pts}-3)}$, where $\chi_0$ is not normalized to the degrees of freedom, $N_{pts}$ is the number of non-zero data points in the $\phi^*_{\pi}$-dependent histogram, and $3$ is the number of free parameters in the fit function Eq.~(\ref{eq:cross02}). The overall averaged $\chi^2$ is $1.25$.

 Table~\ref{tab:sys00} summarizes the systematic uncertainties in this analysis averaged over all accessible kinematic bins shown in Figures~\ref{fig:sft01}-\ref{fig:sft03}. The average total systematic uncertainty is about $11.5\%$. The major sources of the systematic uncertainty are the bin effects, the particle identification (PID) for electron/pion separation and the different physics event generator calculation of the acceptances. The systematic uncertainty for bin effects takes into account bin-size, bin-centering, and different bin number of $\phi^*_{\pi}$ with qudrature sum. The radiative corrections show little systematic dependence on different physics models. 

\begin{figure}[!htb]
\vspace{55mm}
\centering{\includegraphics{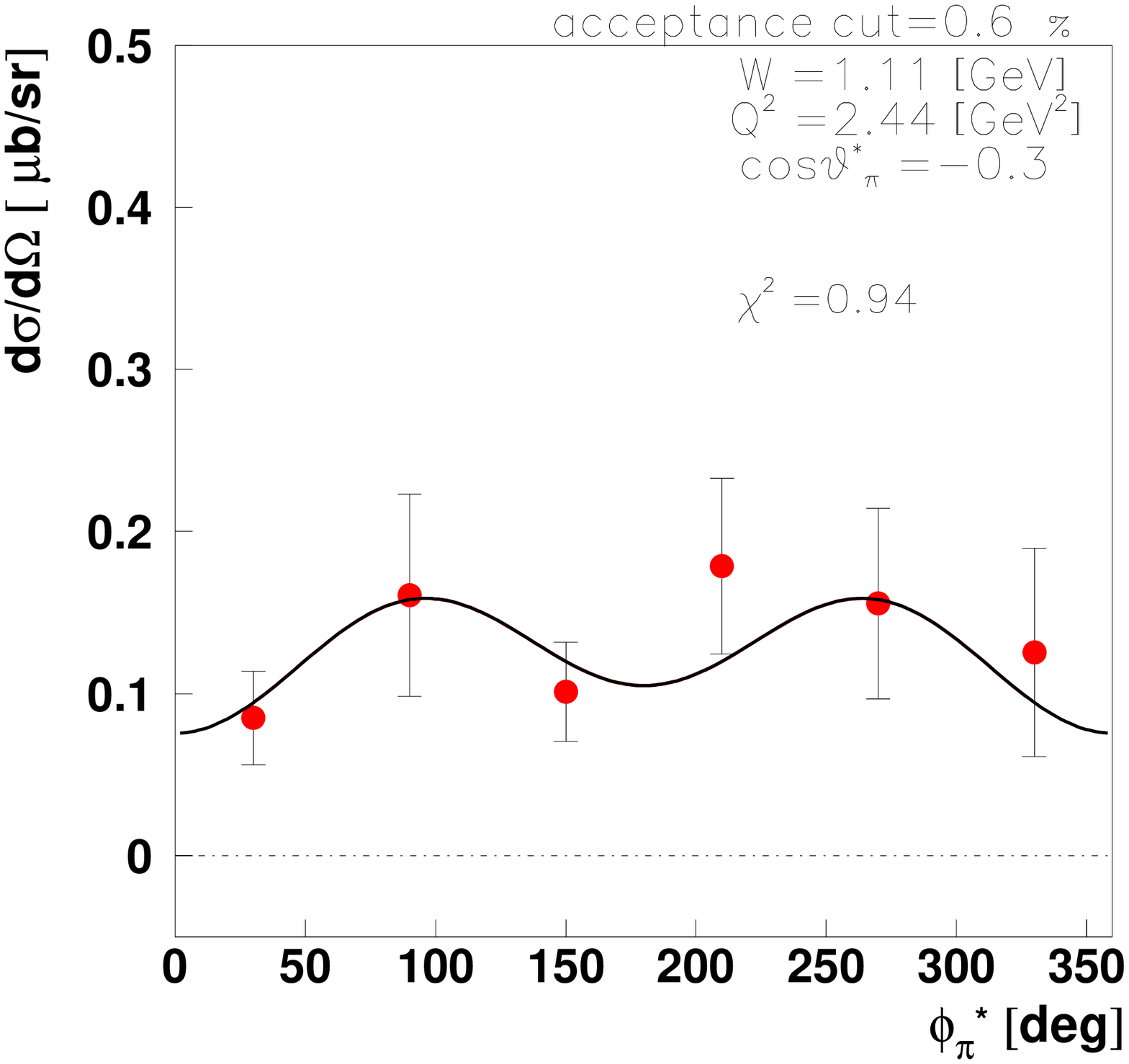}}
\centering{\includegraphics{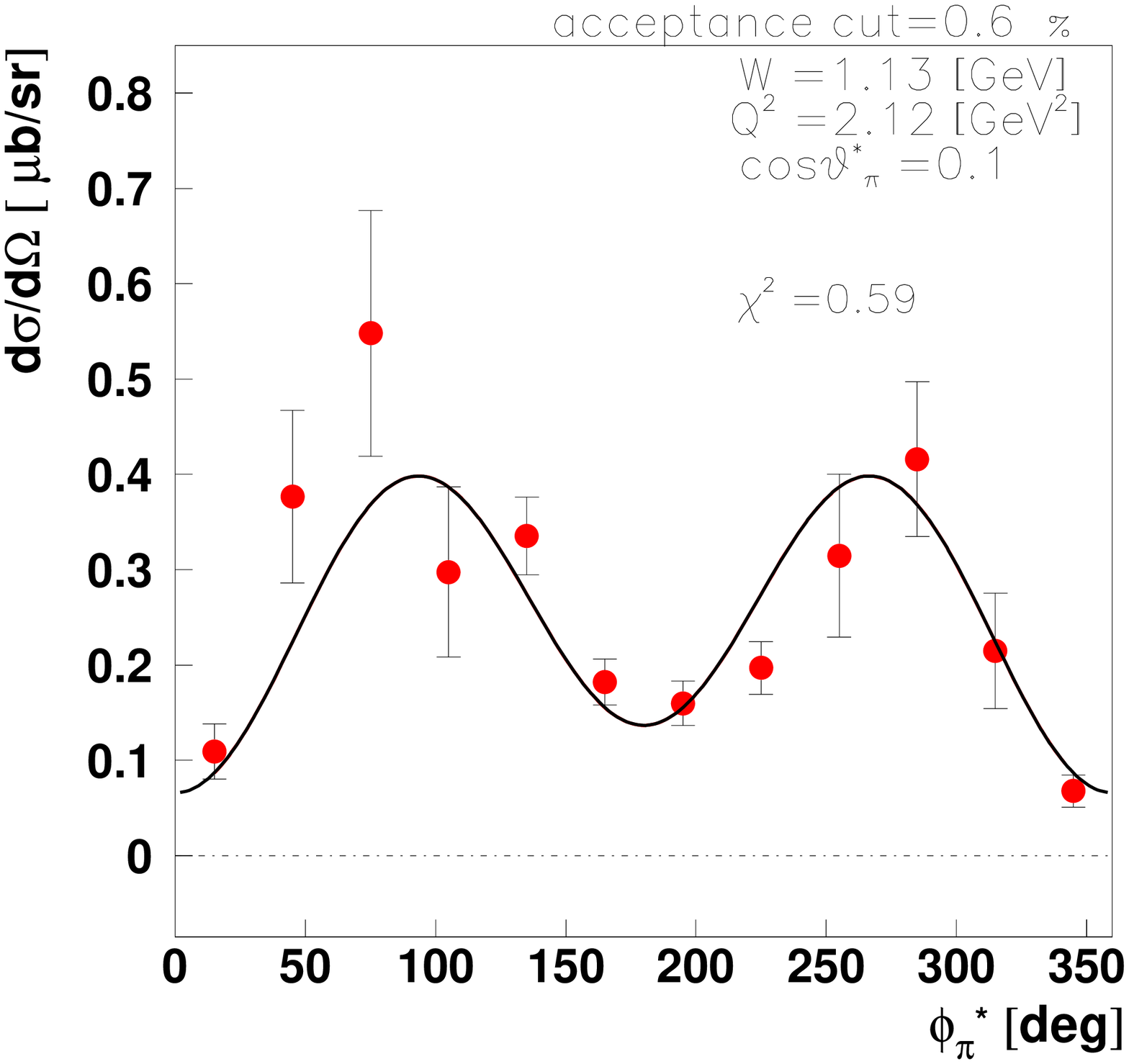}}
        \caption{
        (Color online) Examples of fits to the differential cross sections as function of $\phi^*_{\pi}$ at $W=1.11$, $1.13\;\rm{GeV}$, $Q^2=2.12$, $2.44\;\rm{GeV^2}$, and $\cos\theta_{\pi}^*=-0.3$, $0.1$. The corresponding fit is in~Eq.~(\ref{eq:cross02}).
        }
        \label{fig:crs02}
\end{figure}


\subsection{Extraction of structure functions }
\begin{figure}[!htb]
\vspace{100mm}
\centering{\includegraphics{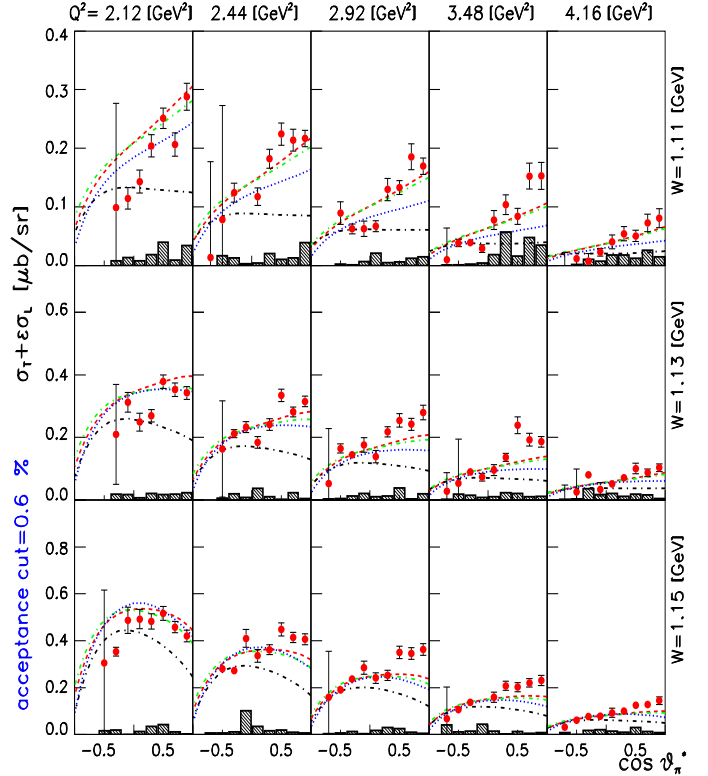}}
        \caption{
        (Color online) The structure function ($\sigma_T+\epsilon \sigma_L$) as a function of $\cos\theta_{\pi}^*$ at $W=1.11-1.15\;\rm{GeV}$ and $Q^2=2.12-4.16\;\rm{GeV^2}$ with model predictions : MAID2003 (red bold dash: full multipoles, green bold dash: without $S_{0+}$, and black dash-dot: without $E_{0+}$) and MAID2007 (blue bold dot). The shaded bars show the estimated systematic uncertainties.
        }
        \label{fig:sft01}
\end{figure}

\begin{figure}[!htb]
\vspace{100mm}
\centering{\includegraphics{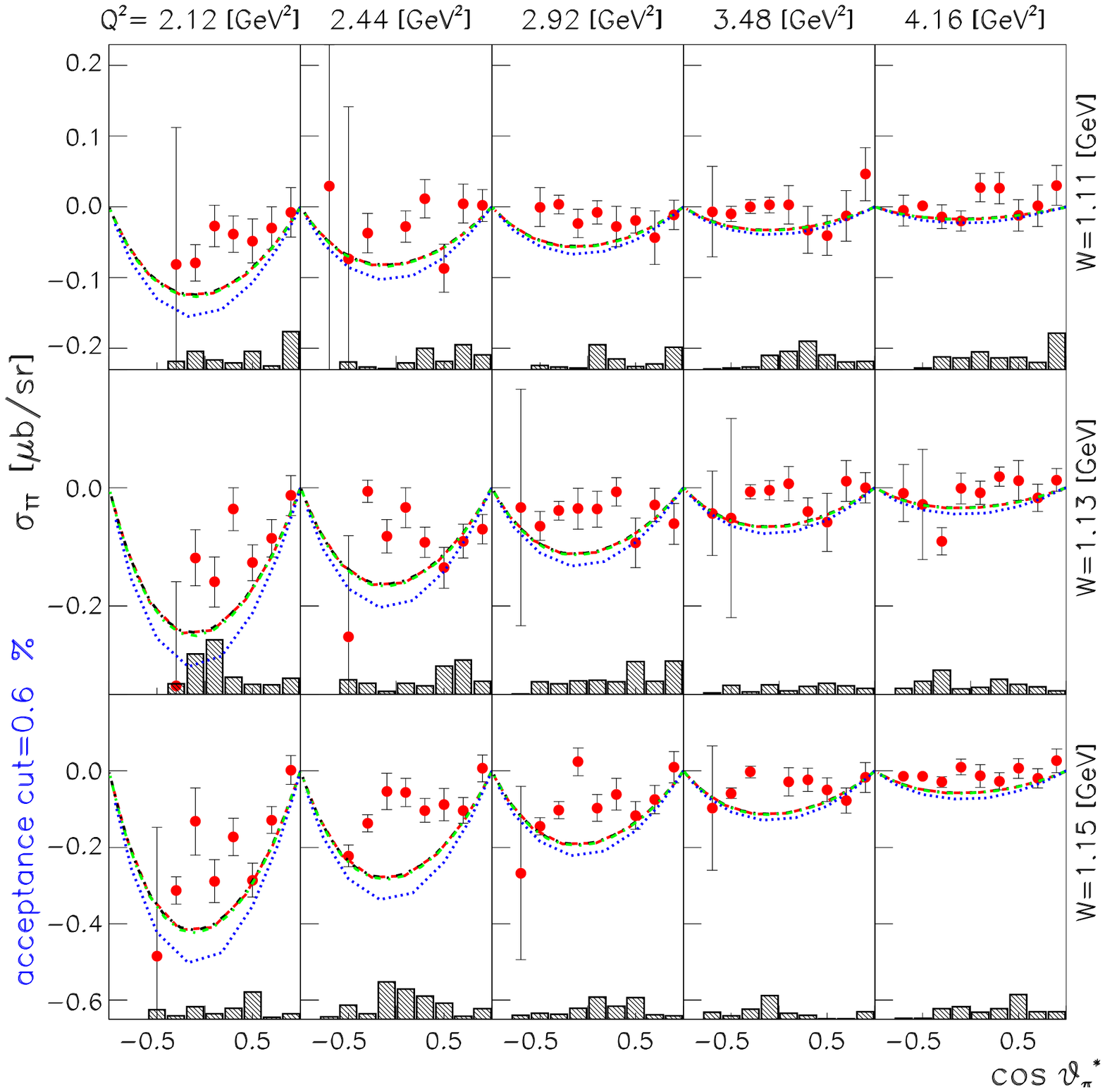}}
        \caption{
        (Color online) The structure function $\sigma_{TT}$ as a function of $\cos\theta_{\pi}^*$ at $W=1.11-1.15\;\rm{GeV}$ and $Q^2=2.12-4.16\;\rm{GeV^2}$ with various model calculations. Curves as in Fig.~\ref{fig:sft01}.
        }
        \label{fig:sft02}
\end{figure}

\begin{figure}[!htb]
\vspace{100mm}
\centering{\includegraphics{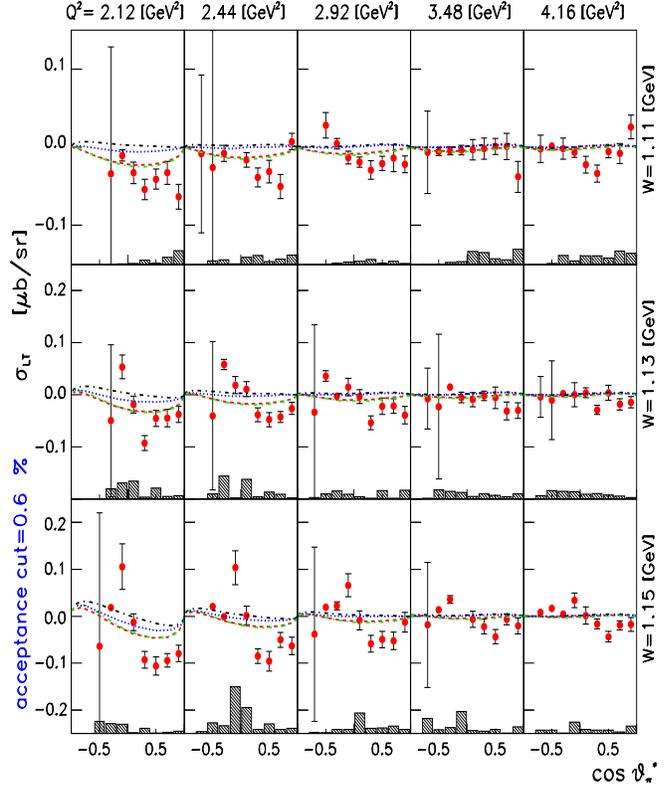}}
        \caption{
        (Color online) The structure function $\sigma_{LT}$ as a function of $\cos\theta_{\pi}^*$ at $W=1.11-1.15\;\rm{GeV}$ and $Q^2=2.12-4.16\;\rm{GeV^2}$ with various model calculations. Curves as in Fig.~\ref{fig:sft01}.
        }
        \label{fig:sft03}
\end{figure}

 The fit of the $\phi^*_{\pi}$-dependent cross sections allows us to access the polarized structure functions. The fitting function has three fit parameters, corresponding to the structure functions $\sigma_{T}+\epsilon\;\sigma_{L}$, $\sigma_{LT}$, and $\sigma_{TT}$: 
\begin{equation}\label{eq:cross02}
\frac{d^2\sigma}{d\Omega^*_{\pi}} = A +  C  \cos 2\phi_{\pi}^* + D  \cos\phi_{\pi}^*~.
\end{equation}
The relation between the structure functions and fit parameters is given by $A = \sigma_{T}+\epsilon\;\sigma_{L}~$, $C =  \epsilon \;\sigma_{TT}~$, and  $D =  \sqrt{2\epsilon(1+\epsilon)} \;\sigma_{LT}~$~\cite{KPark00}.
\begin{table} [!htb]
\begin{center}
\caption{Average systematic uncertainties from various sources for the differential cross sections from this analysis.}
\begin{tabular}{lcc}
\hline
Source & Criterion   & Estimated \\
 &    &  contribution\\
\hline \hline
$e^-$ PID & sampling fraction cut in EC  & $4\%$ \\
  &  ($3 \sigma\to 3.5 \sigma$)   & \\\\
$e^-$ fiducial cut & width ($10\%$ reduced)  &  $2.2\%$\\\\
${\pi}^+$ PID & $\beta$ resolution change  &  $1.3\%$\\
    &  ($2 \sigma_{\rm{TOF}}\to 2.5 \sigma_{\rm{TOF}}$) & \\\\
${\pi}^+$ fiducial cut & width ($10\%$ reduced)  & $3\%$\\\\
missing mass  & neutron missing mass resolution & $1\%$\\
cut              &($3 \sigma_{\rm{MMx}} \to 3.5 \sigma_{\rm{MMx}}$)  &  \\\\
vertex cut & $z$-vertex width ($5\%$ reduced) & $1\%$\\\\
acceptance & event generator dependence  & $4\%$ \\
correction       & AAO\_RAD versus GENEV      &  \\\\
radiative & physics model dependence   & $0.5\%$ \\
correction & Sato-Lee2004 versus MAID2003      &  \\\\
mininum  & aplied $0.6\%$ cut &  \\
acceptance cut & and 6,12 $\phi^*_{\pi}$& $9\%$ \\
and bin effect &  bins     &     \\
\hline
Total  &   &  {$11.5\%$} \\
\hline
\end{tabular}
\label{tab:sys00}
\end{center}
\end{table}
 Figure~\ref{fig:sft01} shows the structure function of $\sigma_T + \epsilon \sigma_L$ as a function of $\cos\theta_{\pi}^*$ in different $Q^2$ bins near the pion threshold region. Figures~\ref{fig:sft02} and~\ref{fig:sft03} show the interference terms $\sigma_{TT}$ and $\sigma_{LT}$, respectively. For the three structure functions shown, several features are notable. First, the $E_{0+}$ multipole in the MAID2003 model plays a dominant role (compare between red bold dash and black dash-dot lines) in both $\sigma_T+\epsilon \sigma_L$ and $\sigma_{TT}$ in the forward angles, and the lower values of $Q^2$ show larger differences between the results with and without $E_{0+}$. Second, the experimental results show a good consistency with the MAID predictions at all $Q^2$. Finally, most of the $\sigma_{TT}$ results particularly at high $Q^2$ are close to zero in the $W=1.11\;\rm{GeV}$ bin, which is the closest to threshold. It is expected that the $d$-wave contribution is absent at threshold, although the MAID calculations still indicate substantial $d$-wave at low $Q^2$.

 The LCSR calculations predict $\phi^*_{\pi}$-independent differential cross sections caused by a complete cancellation of the contribution to $\sigma_{LT}$ and $\sigma_{TT}$ from $G_1$ and $G_2$ (see Section~\ref{gff_lcsr}) at particular values of $Q^2$.  This analysis shows the $\theta_{\pi}^*$ dependence  for the longitudinal-transverse interference term $\sigma_{LT}$, as seen in Fig.~\ref{fig:sft03}. $\sigma_{LT}$ is especially strong for large angles and relatively small $Q^2$. However, this is still quite consistent with the LCSR prediction if we focus on the highest $Q^2$ bin, because the $\phi^*_{\pi}$-independence is only expected for large $Q^2$  in LCSR~\cite{VBraun}. The experimental data for $\sigma_{LT}$ and $\sigma_{TT}$ show small deviations from zero over all $\cos\theta_{\pi}^*$ for all three $W$ bins at $Q^2 = 4.16\;\rm{GeV^2}$, which is possibly caused by the previously mentioned cancellation of the $G_1$ and $G_2$ contributions.

\subsection{Extraction of Legendre moments}
 The Legendre moments for the components of the differential cross sections were extracted by fitting the $\cos\theta_{\pi}^*$ distributions of the $\sigma_T+\epsilon \sigma_L$, $\sigma_{TT}$, and $\sigma_{LT}$ structure functions with first and second order Legendre polynomials. As mentioned in Section IIIa (see Eq.~\ref{eq:stf_legen}), $P_l(\cos\theta_{\pi}^*)$ is the $l ^{th}$-order Legendre polynomial and $D_l^{(T+L)}$, $D_l^{(TT)}$, $D_l^{(LT)}$ are the Legendre moments for $\sigma_T+\epsilon \sigma_L$, $\sigma_{TT}$, and $\sigma_{LT}$, respectively. Each moment can be written as an expansion of magnetic ($M_{l_{\pi^{\pm}}}$), electric ($E_{l_{\pi^{\pm}}}$), and scalar ($S_{l_{\pi^{\pm}}}$) $\pi N$-multipoles or as a factorization of generalized form factors in the light-cone sum-rule framework.  The expansion is truncated at $l_{\pi}=1$, because the $s$-,$p$-wave interference terms, particularly those involving the multipole $E_{0+}$, dominate near the threshold.

\begin{figure}[!htb]
\vspace{95mm}
\centering{\includegraphics{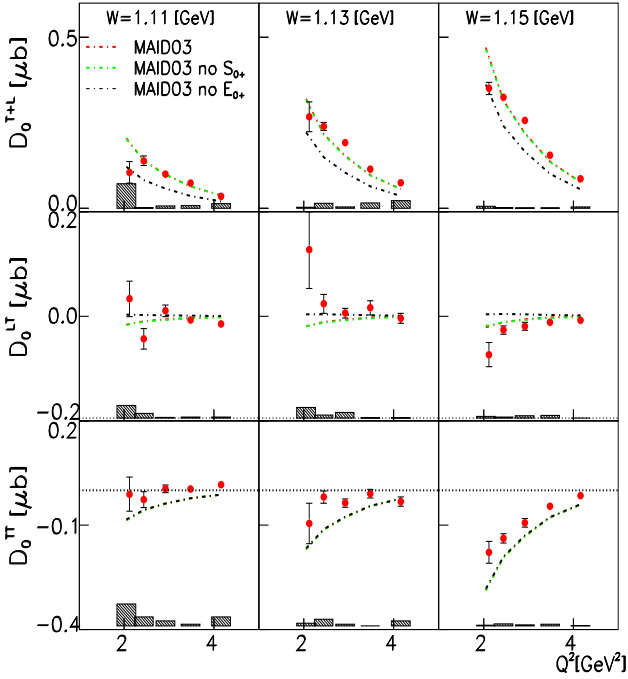}}
        \caption{
        (Color online) $Q^2$ dependence of the Legendre moments for $\sigma_T+\epsilon \sigma_L$, $\sigma_{TT}$, and $\sigma_{LT}$ in the $n\pi^+$ channel at $W=1.11$, $1.13$, and $1.15\;\rm{GeV}$ together with different model predictions:  MAID2003 (red bold dash-dot: full multipoles, green bold dash-dot: without $S_{0+}$, and {black bold dash-dot}: without $E_{0+}$). The shaded bars show the estimated systematic uncertainties.
        }
        \label{fig:legendre00}
\end{figure}

 Figure~\ref{fig:legendre00} shows the $Q^2$-dependent Legendre moments for $\sigma_T+\epsilon \sigma_L$, $\sigma_{TT}$ and $\sigma_{LT}$ at $W=1.11$, $1.13$, and $1.15\;\rm{GeV}$ with different model predictions. Our measurements cover $1.9\;\rm{GeV^2} < Q^2 < 4.5\;\rm{GeV^2}$ with bin center values from $2.12$ to $4.16\;\rm{GeV^2}$. Figure~\ref{fig:legendre00} reveals that the extraction of $D_0^{TT}$ leads to values close to zero over all $Q^2$, particularly in the lowest $W$ bin, whereas the MAID models predict a sizable amplitude.

\subsection{Extraction of the $E_{0+}$ by the LCSR method}
\begin{figure}[!htb]
\vspace{65mm}
\centering{\includegraphics{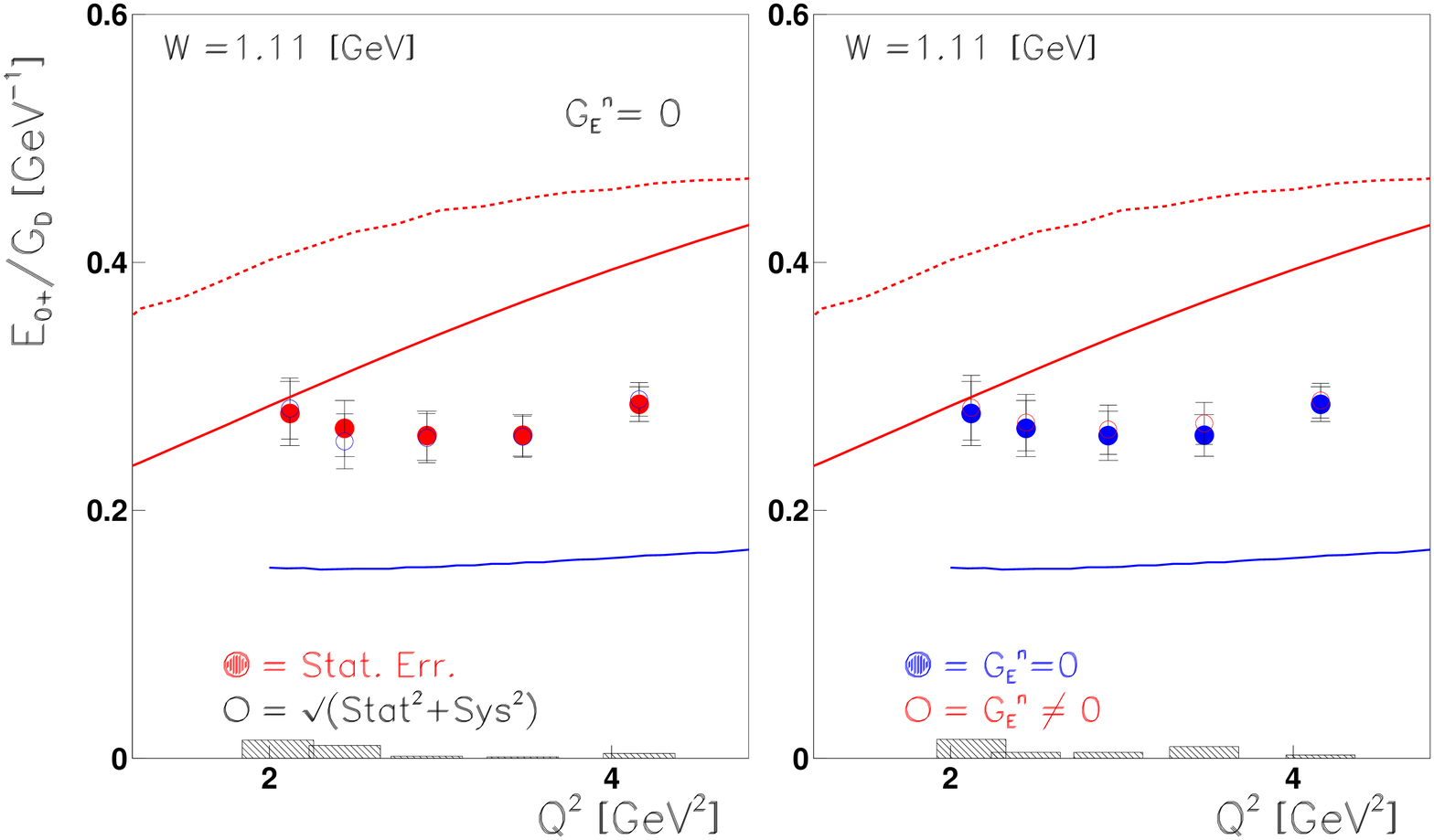}}
        \caption{
       (Color online) The extraction of the $E_{0+}$ multipole divided by the dipole form factor ($G_D$) as function of $Q^2$. The left plot shows the comparison of $E_{0+}$ results taking only statistical or $\sqrt{stat^2+syst^2}$ uncertainties into account. The right plot shows the effect of setting $G_E^n$ to zero. Shaded bars show the systematic errors. Various models are presented, blue solid line: MAID2007 for  $E_{0+}/G_D$, and red solid-dash lines: LCSR (red solid is the LCSR calculation using experimental electromagnetic form factors as input and red dash is pure LCSR)~\cite{VMBraun01}.
        }
        \label{fig:lcrs01}
\end{figure}

The extracted Legendre moments can be directly used in Eq.~(\ref{eq:coeff1}) or Eq.~(\ref{eq:coeff2}) depending on the parametrization of the electric form factor of the neutron ($G_E^n$) and other kinematic constants. Figure~\ref{fig:lcrs01} shows the results of the $Q^2$-dependent $E_{0+}$ multipole divided by the dipole form factor ($G_D$) in the $W$ bin nearest to the threshold. The plot on the left of Fig.~\ref{fig:lcrs01} shows the comparison of  $E_{0+}$ with and without taking systematic uncertainties into account under the assumption of $G_E^n=0$. The plot on the right shows the comparison of $E_{0+}$ multipoles extracted with $G_E^n = 0$ or $G_E^n \ne 0$. MAID2007~\cite{maid2007} and LCSR calculations are also shown.

\begin{figure}[!htb]
\vspace{65mm}
\centering{\includegraphics{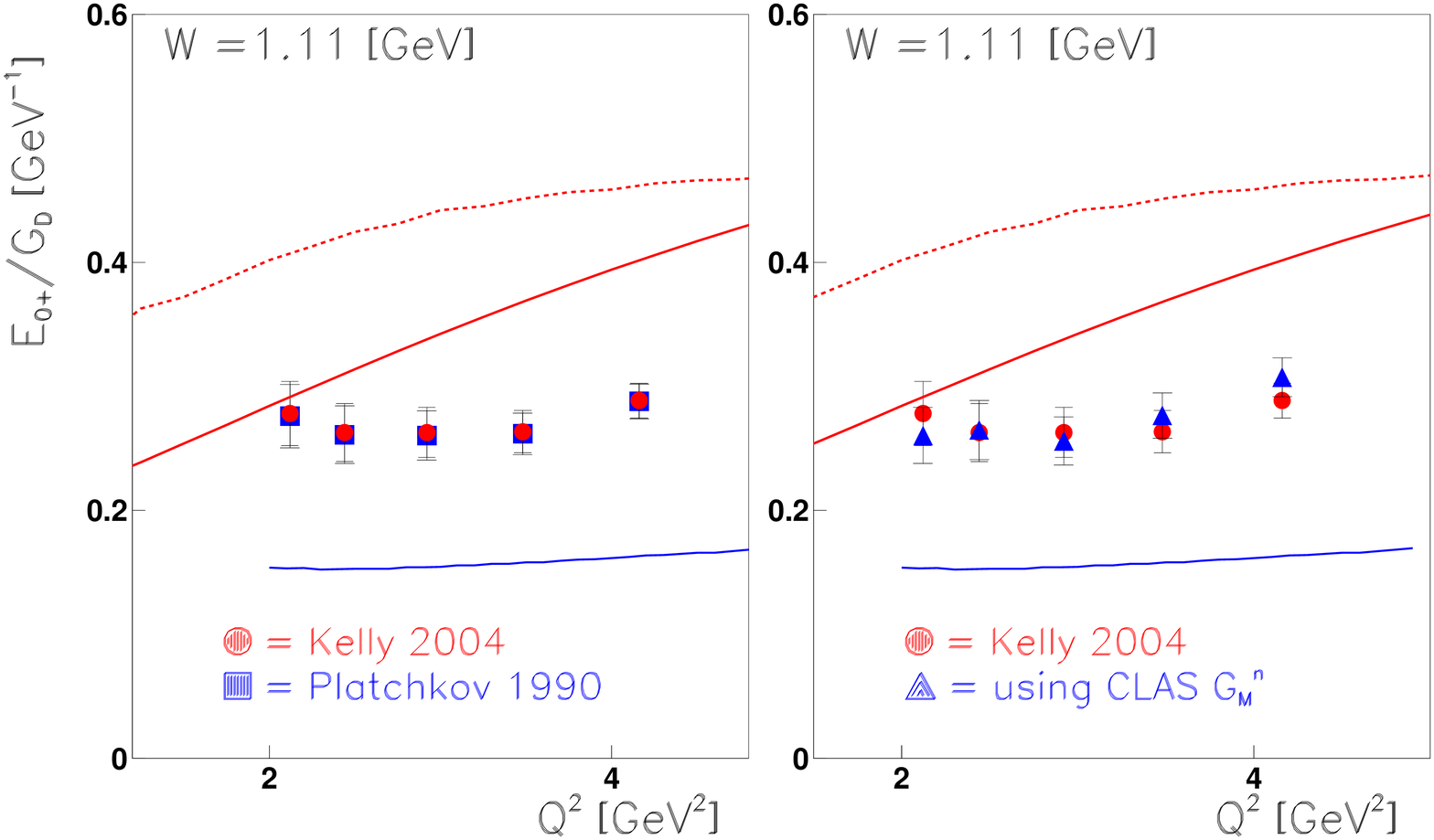}}
        \caption{
        (Color online) The extraction of the normalized $E_{0+}/G_D$ multipole by ignoring the pion mass with different neutron electric form factor parametrizations (left),~blue full squares: Platchkov 1990~\cite{SPlatchkov} and red full circles:~Kelly 2004~\cite{JJKelly}. The right plot shows the  $E_{0+}/G_D$  with different neutron magnetic form factors~blue full triangles:~CLAS measurement~\cite{JLachniet} and red full circles: Kelly 2004~\cite{JJKelly}. Curves as in Fig.~\ref{fig:lcrs01}. 
        }
        \label{fig:nomasspion00}
\end{figure}

 The experimental results for $E_{0+}/G_D $ are about $0.2-0.3\;\rm{GeV^{-1}}$ and almost flat as a function of $Q^2$. The amplitude is larger than predicted by MAID2007 and similar to (or a bit smaller than) the LCSR calculations, although LCSR has a steeper $Q^2$ dependence, which may be caused by the extrapolation in the current LCSR calculations of the form factors to the chiral limit ($m_{\pi}\to0$). The multipole was extracted for two different $G_E^n$ dipole parametrizations~\cite{SPlatchkov,JJKelly} under the assumption of vanishing pion mass. Figure~\ref{fig:nomasspion00} (left) shows that the comparison of both $G_E^n$ parametrizations leads to negligible difference.

 Since CLAS has published the measurement of the magnetic neutron form factor at high momentum transfers with a level of accuracy of $3\%$~\cite{JLachniet}, we can directly substitute this result into our extraction instead of the parametrization. Figure~\ref{fig:nomasspion00} (right) shows the results of $E_{0+}$ at $W=1.11$ GeV for the measured CLAS $G_M^n$ form factor.

%
%
%
%
%
%
\subsection{Extraction of $E_{0+}$ by multipole expansion}
The relations between the invariant amplitudes $f_i$ and the helicity and multipole amplitudes of the cross sections are found in Ref.~\cite{Inna1998Appendix}. Near the pion threshold, the maximum total pion angular momentum taken into account is up to $p$-wave. Therefore, all $f_i$ can be expressed in terms of 
\begin{eqnarray}\label{eq:f_i}
f_1 &=& E_{0+} + 3 \cos\theta_{\pi}^* (E_{1+}+M_{1+})~ ,\nonumber\\
f_2 &=& 2 M_{1+} + M_{1-}~ ,\nonumber\\
f_3 &=& 3 (E_{1+} - M_{1+})~ ,\nonumber~\\
f_4 &=& 0~ ,\nonumber\\
f_5 &=& S_{0+} + 6 \cos\theta_{\pi}^* S_{1+}~ ,~\rm{and}  \nonumber\\
f_6 &=& S_{1-} - 2 S_{1+}~ .\nonumber
\end{eqnarray}
The corresponding helicity amplitudes ($H_i$) are given by
\begin{eqnarray}\label{eq:h_i}
H_1 &=& \frac{-1}{\sqrt{2}} \cos\frac{\theta_{\pi}^*}{2} \sin\theta_{\pi}^* (f_3 + f_4)~ ,\nonumber\\
H_2 &=& -{\sqrt{2}} \cos\frac{\theta_{\pi}^*}{2} (f_1 - f_2 - \sin^2\frac{\theta_{\pi}^*}{2} (f_3 - f_4))~ ,\nonumber\\
H_3 &=& \frac{1}{\sqrt{2}} \sin\frac{\theta_{\pi}^*}{2} \sin\theta_{\pi}^* (f_3 - f_4)~ ,\nonumber\\
H_4 &=& {\sqrt{2}} \sin\frac{\theta_{\pi}^*}{2} (f_1 + f_2 + \cos^2\frac{\theta_{\pi}^*}{2} (f_3 + f_4))~ ,\nonumber\\
H_5 &=& \frac{-\sqrt{Q^2}}{|k_{cm}|} \cos\frac{\theta_{\pi}^*}{2} (f_5 + f_6)~ ,~\rm{and}   \nonumber\\
H_6 &=& \frac{\sqrt{Q^2}}{|k_{cm}|} \sin\frac{\theta_{\pi}^*}{2} (f_5 - f_6)~  .\nonumber
\end{eqnarray}
Here, $k_{cm}$ is the photon momentum in the center-of-mass system.
The structure functions that we measure can be expressed by these helicity amplitudes.
\begin{eqnarray}\label{eq:h_i2}
\sigma_{T}+\epsilon\sigma_{L}&=&\frac{1}{2}\sum_{i=1}^{4}{|H_i|}^2+\epsilon({|H_5|}^2+{|H_6|}^2),\nonumber\\
\sigma_{TT}&=&Re({H_2^*} {H_3} - {H_1^*} {H_4}) ~,~\rm{and}\nonumber\\
\sigma_{LT}&=&\frac{-1}{\sqrt{2}}Re({H_5^*}({H_1} -{H_4}) + {H_6^*}({H_2} + {H_3}))~.\nonumber
\end{eqnarray}

 Since we are interested in the threshold region, we focus on the seven complex multipoles ($E_{0+}$, $E_{1+}$, $M_{1+}$, $M_{1-}$, $S_{0+}$, $S_{1+}$, and $S_{1-}$) with $l \le 1$, and hence 14 quantities must be fit to our data or determined from other data. In particular, the multipoles $M_{1+}^{3/2}$, $E_{1+}^{3/2}$, $S_{1+}^{3/2}$ have large resonance contributions from the $\Delta(1232)$. Their $Q^2$ dependences at the resonance pole are well determined from the CLAS data~\cite{Inna1998Appendix} and are used in this fit, whereas the $W$ dependences are taken from SAID~\cite{said00}.

\begin{figure}[!htb]
\begin{center}
	\includegraphics[angle=0,width=0.35\textwidth]{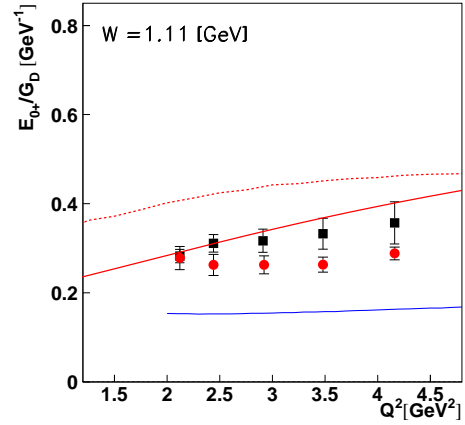}
        \caption{
	  (Color online) 
          \label{fig:multi01} $Q^2$ dependence of $E_{0+}$ normalized by the dipole form factor from the multipole fit. The red bullets are the $E_{0+}/G_D$ results based on LCSR without taking the pion mass into account. The black bullets are the results of multipole. Curves as in Fig.~\ref{fig:lcrs01}.
        }
\end{center}
\end{figure}

 The other multipoles $M_{1-}$, $S_{1-}$ and $M_{1+}^{1/2}$, $E_{1+}^{1/2}$, $S_{1+}^{1/2}$ are fit along with the dominant $E_{0+}$, $S_{0+}$ multipoles by using small start values for $M_{1-}$ and $S_{1-}$. Following this procedure, we are able to fit the measured cross section data and to extract the $E_{0+}$ multipole. Figure~\ref{fig:multi01} shows the results of the $E_{0+}$ multipole in terms of $Q^2$ between $2.0$ and $4.5\;\rm{GeV^2}$ for the lowest $W$ bin near pion threshold.

\begin{figure}[!htb]
\begin{center}
	\includegraphics[angle=0,width=0.34\textwidth]{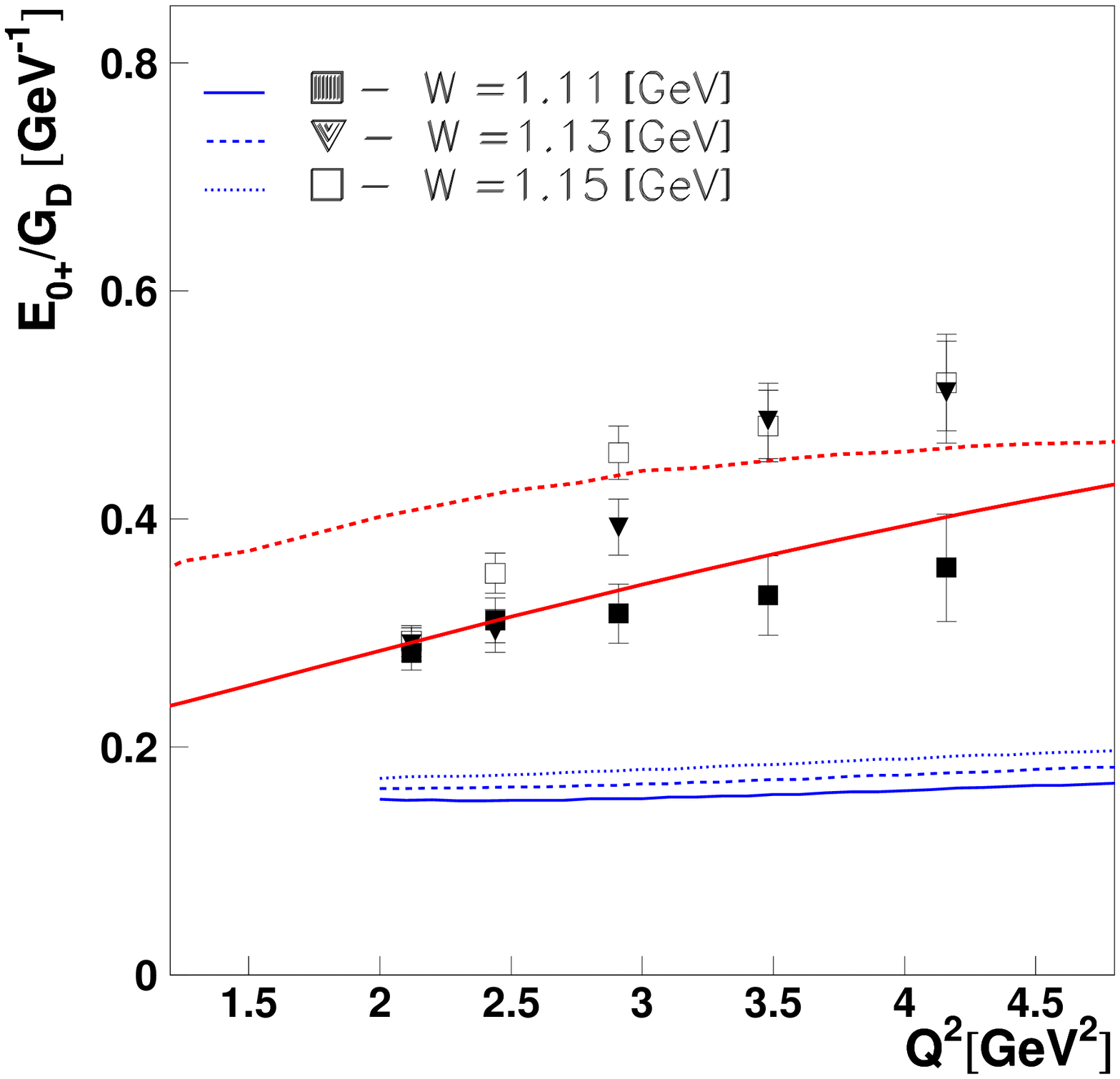}
	\includegraphics[angle=0,width=0.34\textwidth]{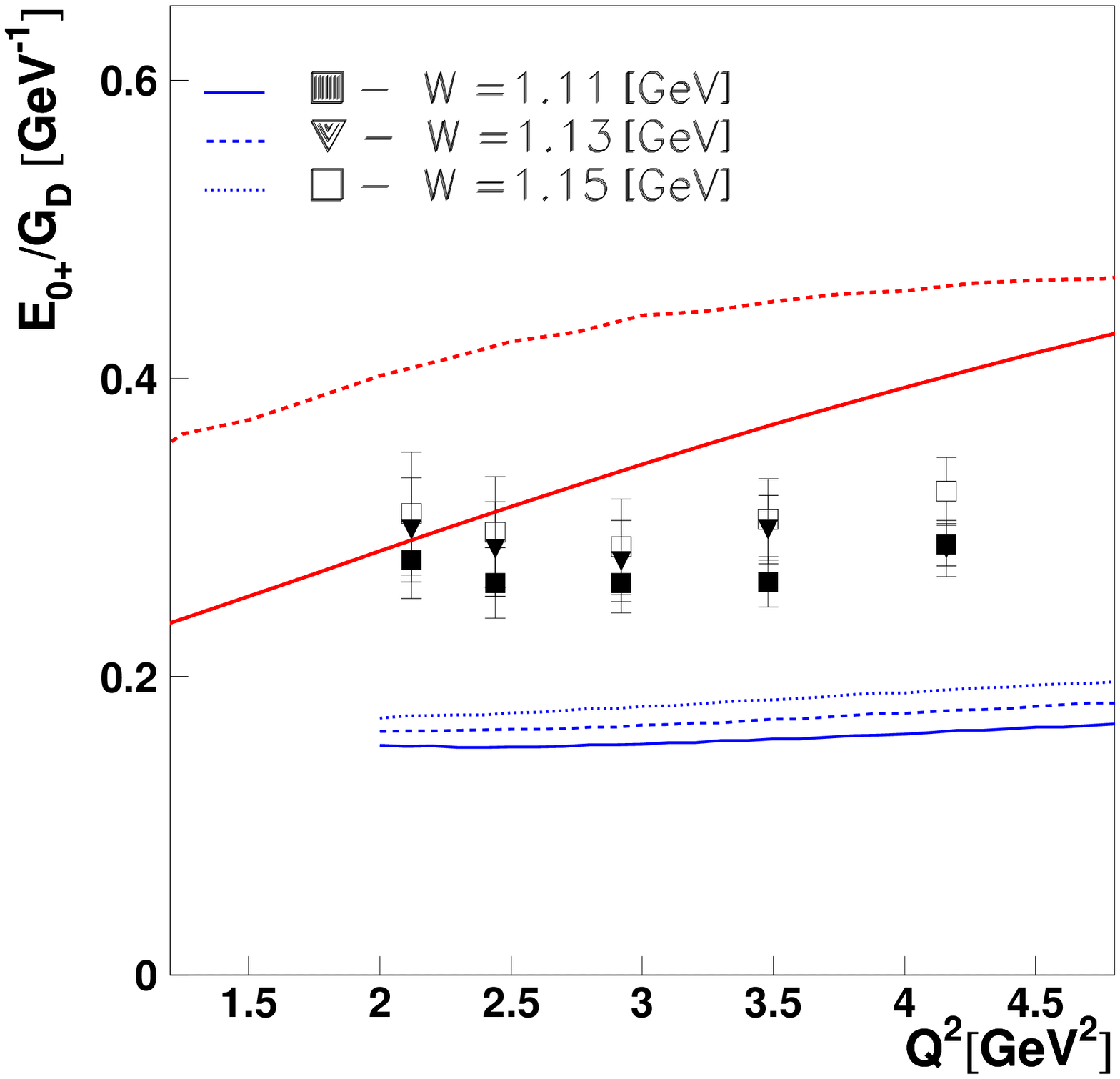}
        \caption{
	  (Color online) 
          \label{fig:multi02} $Q^2$ dependence of $E_{0+}$ normalized by the dipole form factor from the multipole fit (top) and the LCSR (bottom) method (curves) for three $W$ bins.  Curves as in Fig.~\ref{fig:lcrs01}.
        }
\end{center}
\end{figure}

 Figure~\ref{fig:multi02} shows the $Q^2$ dependent $E_{0+}$ multipole extraction from both our multipole fit (top) and the LCSR (bottom) method for three $W$ bins with LCSR~\cite{VBraun} and MAID2007~\cite{maid2007} model. As soon as $W$ is further above the threshold, $s$-wave dominance becomes weaker and resonance and higher order partial waves start to impact the $E_{0+}$ extraction. This expected impact is clearly visible in our multipole fit results and is the reason why in the first place the measurement of the generalized form factors has to be carried out at the pion threshold.

\begin{figure}[!htb]
\begin{center}
	\includegraphics[angle=0,width=0.5\textwidth]{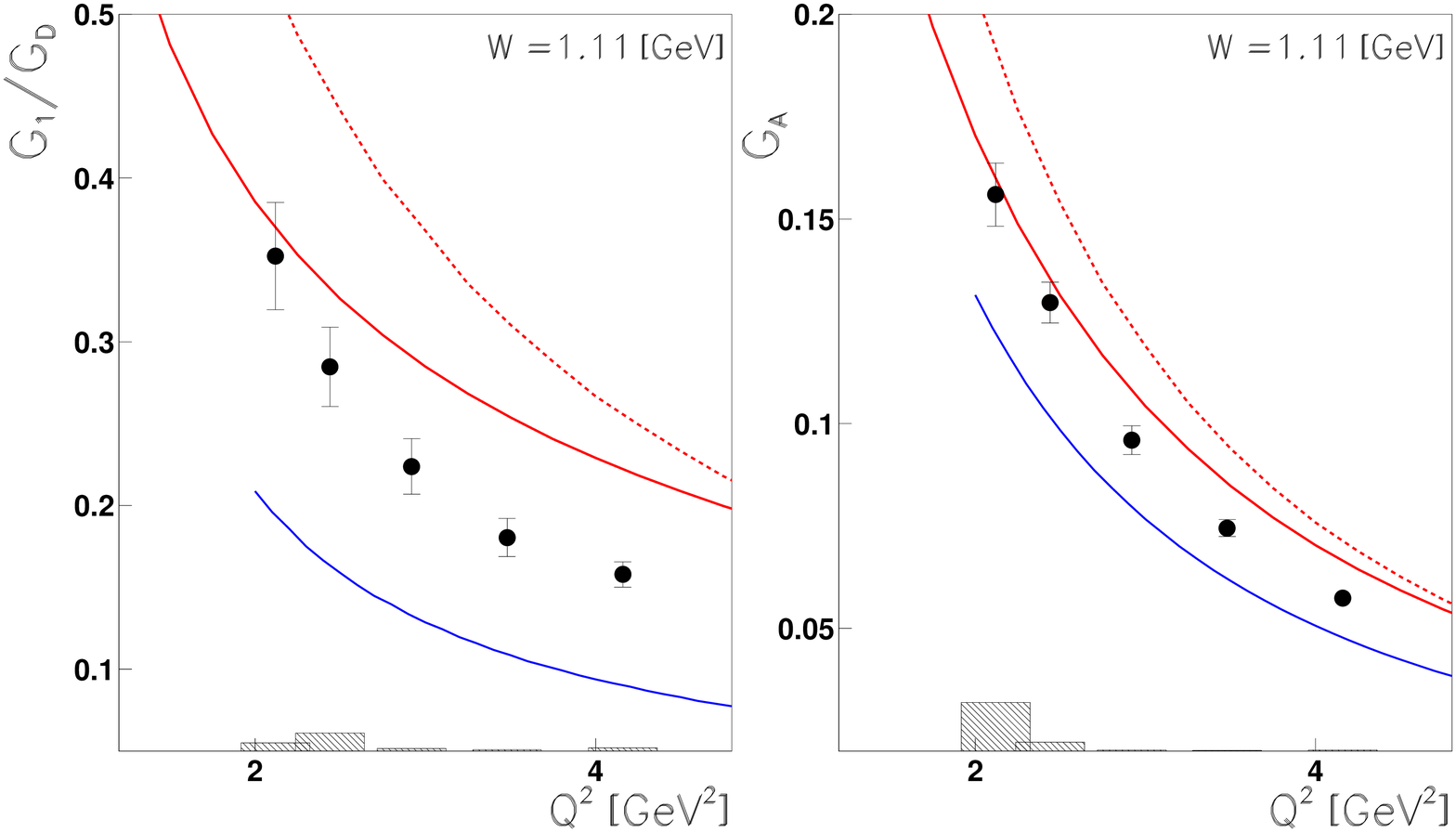}
        \caption{
	  (Color online) 
          \label{fig:formfactor1} $Q^2$ dependence for $n\pi^+$ of $G_1$ normalized by the dipole form factor (left) and axial form factor $G_A$. Curves as in Figure~\ref{fig:lcrs01}.
        }
\end{center}
\end{figure}

\section{Summary}
 We extracted the $E_{0+}$ multipole near pion threshold for $W=1.11 - 1.15\;\rm{GeV}$ at high $Q^2=2.12 - 4.16\;\rm{GeV^2}$ with vanishing and actual pion masses and by taking into account different $G_E^n$ form factor parametrizations. The results for vanishing pion mass show that $E_{0+}/G_D$ is approximately $0.3\;\rm{GeV^{-1}}$ and almost $Q^2$-independent at threshold. This amplitude is larger than the MAID2007 prediction and a little smaller than the LCSR prediction, which has a steeper $Q^2$ dependence. The $Q^2$-independent behavior of the data may be caused by the LCSR method, which is based on the chiral limit $m_{\pi}\to0$. The results from the multipole fit method are consistent with the LCSR method for the lowest $W$ bin.
  Independent of pion mass and $G_E^n$ parametrization considerations, the $n\pi^+$ channel is dominated by the transverse $s$-wave multipole $E_{0+}$. A lack of asymmetry data near the pion threshold does not allow us to extract the generalized form factor $G_2$, but the $E_{0+}$ multipole extraction allows us to obtain $G_1$, and the axial form factor $G_A$ using Eqs.~(\ref{eq:ff0}) and ~(\ref{eq:ff1}). Figure~\ref{fig:formfactor1} shows the $Q^2$-dependent $G_1$ (left) and $G_A$ (right) near-pion threshold. These data give strong constraints on theoretical developments, especially on the extrapolation away from threshold and away from the chiral limit.
\section{Acknowledgement}
We acknowledge the outstanding efforts of the staff of the
Accelerator and the Physics Divisions at Jefferson Lab that
made this experiment possible. This work was supported in
part by the US Department of Energy, the National Science
Foundation, the Italian Istituto Nazionale di Fisica Nucleare,
the French Centre National de la Recherche Scientifique, the
French Commissariat $\grave{a}$ l'Energie Atomique, the United Kingdom's Science and Technology Facilities Council, and the National Research Foundation of Korea.
 The Southeastern Universities
Research Association (SURA) operated the Thomas
Jefferson National Accelerator Facility for the US Department
of Energy under Contract No.DE-AC05-84ER40150.

%

\clearpage

\end{document}